# Asymmetric defect distribution generates bulk Dzyaloshinskii-Moriya interaction in Pt/Co/Pt

Ayaka P. Ohki,[1,†] Tatsuro Karino,[1,†] Daigo Shimizu,[1,†] Takeshi Kato,[1,2] Daiki Oshima,[1,*] and Masahiro Nagao[1,3,*]

[1] *Department of Electronics, Nagoya University, Nagoya 464-8603, Japan.*

[2] *Advanced Measurement Technology Center, Institute of Materials and Systems for Sustainability, Nagoya University, Nagoya 464-8601, Japan.*

[3] *Center for Integrated Research of Future Electronics, Institute of Materials and Systems for Sustainability, Nagoya University, Nagoya 464-8601, Japan.*

[†] *These authors contributed equally: Ayaka P. Ohki, Tatsuro Karino, and Daigo Shimizu.*

[*] *Authors to whom correspondence should be addressed: oshima.daiki.n6@f.mail.nagoya-u.ac.jp and nagao.masahiro@imass.nagoya-u.ac.jp*

## Abstract

The Dzyaloshinskii-Moriya interaction (DMI) in magnetic multilayers plays a pivotal role in spintronic applications by enabling the stabilization and manipulation of chiral spin textures, including chiral domain walls and skyrmions. While interfacial DMI in asymmetric multilayers is well established, a substantial DMI persists in symmetric multilayers, where the opposing interface contributions should, in principle, cancel. The microscopic origin of this effect remains unresolved. Here, we show defect-mediated bulk DMI in sputtered nominally symmetric Pt/Co/Pt multilayers. Using Lorentz transmission electron microscopy and aberration-corrected scanning transmission electron microscopy, we systematically elucidate the effect of argon sputter gas pressure on both magnetic and structural properties. Increasing the pressure markedly enhances the net DMI, thereby stabilizing chiral spin textures. Atomic-scale analysis reveals the absence of well-defined interfaces, confirming that the DMI predominantly arises in the bulk rather than at the interfaces. However, macroscopic compositional asymmetries are remarkably weak, and spin-orbit coupling shows no significant enhancement, rendering them insufficient to account for the observed DMI. Crucially, our statistical analysis of dislocations uncovers a strongly biased spatial distribution. These asymmetrically distributed dislocations, accompanied by local atomic displacements, combined with compositional gradients, generate the bulk DMI through the three-site Fert-Levy mechanism. Our results establish defect engineering as an effective strategy for controlling DMI, offering new design principles for spintronic devices.

## Main text

The interfacial Dzyaloshinskii-Moriya interaction (DMI) in magnetic multilayers is a cornerstone of modern spintronics, facilitating the stabilization of chiral spin textures such as chiral domain walls (DWs)[1,2] and skyrmions.[3–6] These chiral spin textures display remarkable properties, including current-driven dynamics,[1–4,7–9] deterministic switching,[10,11] and nonreciprocal spin wave propagation,[12,13] making them promising for next-generation memory, logic devices and neuromorphic computing devices. The DMI typically originates from inversion symmetry breaking at interfaces between magnetic layers and nonmagnetic heavy metals possessing strong spin-orbit coupling (SOC).[14–21] However, a puzzling contradiction to this interfacial framework has emerged: a substantial DMI is observed in nominally symmetric multilayers such as sputtered Pt/Co/Pt[22,23] and Pd/Co/Pd,[24–26] where both interfaces should, in principle, contribute equally but with opposite signs, theoretically canceling out. Unraveling the microscopic origin of this "hidden" DMI is essential for advancing both fundamental understanding and device optimization.

Recent studies have shown that sputtering conditions—including argon gas pressure,[27] substrate temperature,[28] and ion irradiation[29]—significantly influence the DMI in Pt/Co/Pt, initially attributed to interfacial intermixing.[27,28] However, theoretical calculations predict robustness[30] or even enhancement[31] of the DMI with intermixing, contradicting this simple interpretation. Meanwhile, macroscopic compositional gradients have been identified as a source of bulk DMI,[32–35] breaking inversion symmetry within films. Despite these advances, quantitative analysis of compositional asymmetries in sputtered multilayers remains limited.[36] Additionally, recent demonstrations of defect-induced DMI in centrosymmetric bulk materials[37–40] suggest that structural imperfections may play a critical role in the DMI. However, the lack of atomic-scale characterization in polycrystalline, ultrathin films has prevented a comprehensive understanding of how composition, defects, and DMI are interconnected.

Here, we employ Lorentz transmission electron microscopy (LTEM) and aberration-corrected high-angle annular dark-field scanning transmission electron microscopy (HAADF-STEM) to elucidate the microscopic origin of DMI in sputtered nominally symmetric Pt/Co/Pt multilayers. We demonstrate that increasing argon sputter gas pressure significantly enhances the net DMI, despite the absence of well-defined interfaces. While compositional gradients exist, their macroscopic asymmetries are too weak to account for the observed DMI. Through statistical analysis, we reveal a strongly biased dislocation distribution within the compositional gradient regions. This asymmetry, combined with compositional gradients, generates substantial DMI through atomic displacements in the three-site Fert-Levy framework.[41] Our findings demonstrate that defect distributions can serve as a key microscopic factor governing the emergence of DMI in sputtered multilayers and provide new insights into spin-orbit physics in technologically relevant thin films.

We fabricated multilayers of substrates/Ta(5)/$[Pt(1)/Co(1)/Pt(1)]_{15}$/Pt(3) by room-temperature magnetron sputtering at argon pressures $P$ = 0.4, 1.0, 1.5, and 2.0 Pa (hereafter referred to as the 0.4-, 1.0-, 1.5-, and 2.0-samples). The numbers in parentheses denote thickness in nanometers. Unlike previous studies employing pressure gradients during deposition to induce asymmetry,[27] we maintained a constant pressure. Detailed deposition conditions, including the sputtering system geometry, target-to-substrate distance, and power and voltage parameters for each layer and each argon pressure, are provided in Section VI of the supplementary material. We note that the notation $[Pt(1)/Co(1)/Pt(1)]_{15}$ reflects the cross-sectional structural perspective revealed by HAADF-STEM, in which each Co-rich region appears sandwiched between Pt-rich regions above and below. In practice, the deposition sequence was ...Pt(2)/Co(1)/Pt(2)/Co(1)..., where successive 2 nm Pt layers are each shared between adjacent Co layers, yielding an effective 1 nm Pt environment above and below each Co layer as observed in cross-section. The sputtering shutter was not closed within the Pt layers during deposition. Multilayers were deposited simultaneously on $Si_3N_4$ membranes for LTEM and on Si/$SiO_2$ for HAADF-STEM and magnetic measurements. In high-resolution HAADF-STEM imaging, to minimize projection artifacts, we selected regions where the specimen thickness was smaller than the grain diameter ($L \approx 20$ nm for the 0.4-sample), thereby reducing the likelihood of grain overlap along the electron beam direction. Importantly, the contrast features we observe at atomic resolution are qualitatively distinct from those expected for projection of a perfectly sharp but dome-shaped interface. Whereas such a projection would produce a smoothly blurred intensity profile, we instead observe bending and vertical splitting of individual atomic monolayers — indicative of genuine single-monolayer deflections along the beam direction — which cannot be attributed to a geometrical projection effect. Regions exhibiting pronounced curvature of atomic planes that could introduce projection ambiguity were excluded from the quantitative analyses in Figs. 3 and 4. Furthermore, the absence of sharp interfaces was confirmed across the entire statistical dataset (114 and 122 regions for the 0.4- and 1.0-samples, respectively; see Tables S4 and S5), making a systematic projection artifact an unlikely explanation. For the $P \geq 1.0$ Pa samples, the grain sizes ($L \approx 5$–$10$ nm) were too small to exclude grain overlap along the beam direction reliably; atomic-resolution analysis was therefore not performed on these samples.

We observe enhanced DMI with increasing sputter pressure. Figure 1 displays LTEM images captured at sample tilt angles $\alpha = 0°$ and 30°. The 0.4-sample exhibits pronounced DW contrasts at $\alpha = 0°$ [Fig. 1(a), top], characteristic of Bloch-type DWs, which indicates that the DMI is too weak to form chiral DWs. In contrast, the samples deposited at $P \geq 1.0$ Pa show no magnetic contrast at $\alpha = 0°$ [Figs. 1(b)–1(d), top] but clear contrast at $\alpha = 30°$ [Figs. 1(b)–1(d), bottom], unambiguously indicating Néel-type DWs and skyrmions.[24] We note that the observation of Néel-type spin textures in place of the magnetostatically preferred Bloch-type DWs is itself direct evidence for the presence of a substantial DMI: as established in the literature,[1,42] the DMI must exceed the critical value $|D_c|$ to overcome the magnetostatic energy penalty and stabilize Néel-type DWs over Bloch-type DWs. The transition from Bloch-type DWs in the 0.4-sample to Néel-type DWs and skyrmions in the $P \geq 1.0$ Pa samples therefore constitutes a qualitative and model-independent demonstration of enhanced DMI with increasing sputtering pressure, independent of any quantitative estimation of $|D_c|$. Our recent observation of field-free skyrmions in Pt/Co/Ta deposited at $P = 2.0$ Pa[43] further corroborates enhanced DMI under high-pressure conditions. We note that all LTEM images in Fig. 1 were acquired in the as-grown state, without any prior application of an external magnetic field. The spontaneous stabilization of skyrmions under field-free conditions in the $P \geq 1.0$ samples is consistent with the presence of a sufficiently large DMI: as demonstrated by our recent micromagnetic simulations[43], magnetic multilayers possessing a large DMI ($\geq 0.6$ mJ/m$^2$) proportional to the magnetic anisotropy and dipolar energies can spontaneously form skyrmions in the as-grown demagnetized state, without requiring an applied field to nucleate or stabilize them.

We note that the absence of LTEM contrast at $\alpha = 0°$ in the $P \geq 1.0$ Pa samples does not, in itself, constitute direct evidence for a uniform Néel-type spin texture throughout the multilayer thickness. As demonstrated by Legrand et al.,[44] magnetic multilayers with perpendicular anisotropy generally stabilize hybrid chiral DW structures, in which the DW magnetization is continuously twisted from a counterclockwise Néel orientation defined by the DMI in the bottommost layers, through a Bloch-type orientation in the intermediate layers, to a clockwise Néel orientation in the topmost layers, as a result of competition between the DMI and interlayer dipolar interactions. To investigate whether such hybrid chiral structures are formed in our system, we performed micromagnetic simulations[45] of the DW profile in the full multilayer geometry including all magnetic Co layers and nonmagnetic Pt spacers, following the methodology of Legrand et al.[44] The results, presented in Fig. S3 and Table S2 of the supplementary material, confirm that hybrid chiral DW structures are stabilized over most of the investigated DMI range, with no evidence for purely uniform Néel-type DWs. To connect these simulation results quantitatively to our LTEM observations, we evaluated the thickness-averaged Bloch component, defined as $\langle \sin\varphi \rangle = (1/N)\Sigma \sin\varphi(z_i)$, where $\varphi(z_i)$ is the azimuthal angle of the DW magnetization in the $i$-th Co layer and $N = 15$ is the total number of Co layers. This quantity is proportional to the integrated electron phase shift responsible for LTEM contrast at $\alpha = 0°$, and thus provides a direct measure of the expected signal intensity: $\langle \sin\varphi \rangle = 1$ corresponds to a purely Bloch-type DW with maximum contrast, whereas $\langle \sin\varphi \rangle = 0$ corresponds to a purely Néel-type configuration with negligible contrast. For the 0.4-sample, with a DMI comparable to that reported for unstrained Pt/Co/Pt by Gusev et al.[46] ($D \approx 0.1$–$0.4$ mJ/m$^2$), $\langle \sin\varphi \rangle$ is estimated to be approximately 0.45–0.59, reflecting a substantial Bloch component distributed through the multilayer and consistent with the observed Bloch-type contrast at $\alpha = 0°$ [top panel of Fig. 1(a)]. As the sputter pressure increases, the net DMI is enhanced and $<\sin\varphi>$ decreases systematically. Once $<\sin\varphi>$ falls below the detection threshold of our LTEM setup, the $\alpha = 0°$ contrast disappears [top panels of Fig. 1(b)–(d)], even though the spin texture remains of hybrid chiral character. The observed transition from pronounced Bloch-type contrast at $\alpha = 0°$ in the 0.4-sample to its absence in the $P \geq 1.0$ Pa samples therefore reflects the progressive suppression of the Bloch component within the hybrid chiral DW structure, rather than a transition to a uniform Néel-type configuration, and the DMI-driven suppression of $<\sin\varphi>$ constitutes qualitative evidence for enhanced DMI with increasing sputter pressure.

Previous work on Pt/Co/Pt reported a transition from intermixing of Pt and Co to well-defined interfaces at $P = 0.8$ Pa,[27] yet direct atomic-scale observation was lacking. Therefore, we use HAADF-STEM, which exhibits brightness (contrast) that is approximately proportional to the square of the atomic number $Z$, to elucidate interface and intermixing at the atomic scale for the 0.4- and 1.0-samples. Figures 2(a) and 2(d) show cross-sectional HAADF-STEM images. The 0.4-sample displays relatively flat layers with grain sizes of $L \approx 20$ nm, while the 1.0-sample exhibits composition bending (though lattice planes remain relatively flat) and smaller grains ($L \approx 5$–$10$ nm). Both samples show mosaic crystal structures within grains. Crucially, neither sample possesses well-defined interfaces. We observe compositional gradients, as evidenced by the intensity profiles in Figs. 2(c) and 2(f) obtained from the rectangular regions in Figs. 2(b) and 2(e). These profiles display sinusoidal-like modulations, contrasting with the square waves expected for sharp interfaces.[47] The sharp peaks correspond to atomic columns and lattice fringes in the 0.4- and 1.0-sampes, respectively. Some broad peaks in the C-D profile of Fig. 2(f) exhibit asymmetric shapes. This asymmetry does not indicate compositional asymmetry but is caused by partial atomic disorders resulting in the low signal-to-noise ratio and the low intensity of the sharp peaks. Note that both samples exhibit greater intermixing near grain boundaries, with decreasing intermixing away from these areas (see Fig. S1 in the supplementary material). The absence of well-defined, sharp interfaces, with compositional intermixing extending over a characteristic length scale of approximately 1 nm, as evidenced by the sinusoidal-like intensity profiles, indicates that the conventional picture of interfacial DMI arising from atomically sharp heavy-metal/ferromagnet boundaries is not applicable here, and that the observed DMI instead predominantly originates from the bulk of the film rather than from well-defined interfaces.

Within the Fert-Levy framework,[41] the DMI enhancement could arise from increased compositional asymmetry, enhanced effective SOC, or atomic displacements. We systematically examine each factor. To quantify compositional asymmetry, we analyzed intensity profiles from 114 and 122 regions for the 0.4- and 1.0-samples, respectively, measuring gradient distances $d_u$ and $d_l$ from the Co-richest atomic layers to the Pt-richest atomic layers above and below, respectively [the insets of Figs. 3(a) and 3(b)] (see Tables S4 and S5 in the supplementary material). For the 1.0-sample, small polycrystalline grains necessitated the lattice fringes rather than atomic columns. We frequently observe two adjacent Co-richest (Pt-richest) atomic layers above and below in the 1.0-sample, indicating a shorter interdiffusion distance compared to the 0.4-sample; in such cases, we adopted the closest distance between the Co-richest and Pt-richest atomic layers. Additionally, we excluded regions with considerable Pt diffusion. Figures 3(a) and 3(b) show substantial dispersion in $d_u$ and $d_l$ values. The averages are $d_u$(ave) = 1.688 nm, $d_l$(ave) = 1.569 nm, and the differential $d_{l-u}$(ave) = −0.119 nm for the 0.4-sample, and $d_u$(ave) = 1.427 nm, $d_l$(ave) = 1.388 nm, and $d_{l-u}$(ave) = −0.039 nm for the 1.0-sample. These results reveal weak macroscopic compositional asymmetries, with the 1.0-sample showing even weaker asymmetry than the 0.4-sample despite its stronger DMI.

Following the methodology of Ref. 35, in which the bulk DMI constant $D$ scales linearly with the composition gradient $\Delta x/t$, where $\Delta x = x_s - x_e$ is the composition difference between the starting and ending layers and $t$ is the thickness over which the gradient extends, we estimate the net bulk DMI energy $D_{net}$ from the measured gradient distances $d_u$ and $d_l$. Specifically, $d_u$ and $d_l$, defined as the distances from the Co-richest atomic layer to the upper Pt-richest and lower Pt-richest atomic layers, respectively, play the role of $t$ in Ref. 35 for the upper (Pt/Co) and lower (Co/Pt) compositional gradient regions, respectively. The DMI contributions from the upper and lower gradient regions are estimated separately, and $D_{net}$ is obtained from their difference, which is nonzero when $d_u \neq d_l$. To obtain a conservative estimate, we assume the maximum possible composition gradient in each region, i.e., $\Delta x = |x_s - x_e| = 1$, corresponding to a composition change from 100 % Pt to 100 % Co (or vice versa) across the gradient distance. This assumption deliberately overestimates the actual composition gradient in each region, and the resulting $D_{net}$ values should therefore be interpreted as conservative upper bounds. Figures 3(c) and 3(d) present $D_{net}$ dispersions corresponding to Figs. 3(a) and 3(b), showing large variations up to $|D_{net}| \approx 0.6$ mJ/m$^2$ due to the dispersion in $d_u$ and $d_l$. However, the average values are $|D_{net}(\text{ave})| \approx 0.05$ mJ/m$^2$ and $\approx$ 0.03 mJ/m$^2$ for the 0.4- and 1.0-samples, respectively. Notably, $D_{net}$(ave) for the 1.0-sample is far too small to explain the LTEM observations and is not significantly different from the 0.4-sample. Moreover, $D_{net}$(ave) of the 0.4-sample is an order of magnitude smaller than previously reported values.[46]

We also examined SOC through electron energy loss spectroscopy, which probes the same dipole-allowed core-to-unoccupied-state transitions as X-ray absorption spectroscopy (XAS) and thereby provides equivalent spectroscopic information at the Co $L_{2,3}$ edges.[48–51] By quantitatively evaluating the branching ratio $B = I(L_3)/[I(L_3) + I(L_2)]$, which is theoretically given by $B = B_0 + \langle l \cdot s/n_h \rangle$, where $B_0$ is the branching ratio without SOC, $l$ and $s$ are the orbital and spin angular momenta, and $n_h$ is the number of $d$-holes,[52] enhanced SOC in Co-based multilayers has been reported to yield $B$ = 0.75–0.82, significantly above the ideal statistical value of $B \approx 0.667$ corresponding to an $I(L_3)/I(L_2)$ ratio of 2:1.[46-49] We obtained $B$ = 0.66 for the 0.4-sample and $B$ = 0.65 for the 1.0-sample from the whole-film regions, and $B$ = 0.65 from the grain boundary region of the 0.4-sample — all close to the ideal statistical ratio and well below the enhanced values reported for systems with substantial SOC. The near-identical branching ratios of the 0.4- and 1.0-samples, despite their markedly different DMI strengths, confirm that SOC enhancement does not account for the observed difference in DMI. Further details are provided in Section III of the supplementary material.

With compositional asymmetry and the SOC enhancement ruled out, we focus on atomic displacements induced by defects. It is well known that defect density increases with sputter gas pressure,[27,53] suggesting a direct connection to the DMI. We therefore conduct detailed HAADF-STEM analysis of structural defects within grains. Highly disordered grain boundaries make it challenging to identify asymmetry, and their vicinity with considerable Pt diffusion (see Fig. S1 in the supplementary material) generates negligible magnetization. Therefore, we focus on dislocations within the grains. We selected the 0.4-sample for statistical analysis due to its relatively larger grains ($L \approx 20$ nm) facilitating reliable identification with atomic resolution, noting that substantial DMI in sputtered Pt/Co/Pt with $P \leq 0.4$ Pa is well established.[22,23,27,28,46] In the $P \geq 1.0$ samples, the grain sizes are too small ($L \approx$ 5–10 nm) to reliably isolate individual dislocations with atomic resolution, and grain overlap along the electron beam direction in cross-sectional specimens cannot be excluded, rendering a statistically reliable dislocation analysis infeasible with the current specimens. Extending this analysis to higher-pressure samples will require advances in specimen preparation techniques capable of producing TEM lamellae thinner than the fine grain size of $L \approx$ 5–10 nm, as well as potentially alternative characterization approaches such as atom probe tomography or three-dimensional electron tomography. Figure 4(a) shows a representative Fourier-filtered HAADF-STEM image, revealing streak-like blurred contrasts between atomic columns (green arrows), primarily parallel to the layers. Intensity profiles along lines 1-1′ to 4-4′ in Fig. 4(b) confirm these streak-like contrasts, which can be identified as screw [Fig. 4(c)], edge [Fig. 4(d)], or mixed dislocations. Note that due to the small grain size and high degree of crystalline disorder, we could not apply the Burgers circuit, which identifies dislocation types by tracing the path around the fully crystalline region around a dislocation. The 5-5′ line profile exemplifies streak-contrast-free regions.

We analyze the positional distribution of 241 dislocations relative to the Co-richest atomic layers (see Table S56 in the supplementary material). As shown in Fig. 4(a) (right panel), intensity profiles combined with HAADF-STEM images simultaneously identify the atomic layer positions containing dislocation ($N_d$) and the Co-richest layers ($N_{Co}$). Counting layers from the bottom Pt-richest atomic layer and excluding dislocations within three atomic layers from the top and bottom Pt-richest layers (presumed non-magnetic[35]), we determined the atomic layer differences $N_{d\text{-}Co} = N_d - N_{Co}$. Figure 4(e) reveals a striking asymmetry: 51.8% of dislocations lie above the Co-richest atomic layer (positive $N_{d\text{-}Co}$), 26.6% below (negative), and 21.6% coincide (zero). This biased distribution—nearly twice as many dislocations above versus below the Co-richest atomic layers—provides direct evidence for inversion symmetry breaking. This preferential positioning of dislocations likely arises from the lattice mismatch effects analogous to those in single-crystal Fe/MgO/Fe magnetic tunnel junctions,[54] where the lattice mismatch between the MgO growth layer with the larger lattice constant and the Fe underlayer with the smaller one promotes dislocation formation. In this situation, Pt corresponds to the growth layers and Co to the underlayers in Pt/Co/Pt.

The three-site Fert-Levy model[41] describes the DMI vector between neighboring magnetic atoms mediated by surrounding heavy atoms through three vectors forming a triangle and the heavy metal atom's position angle [Fig. 4(f)]. Atomic displacements modify this DMI vector. A recent study has demonstrated that uniaxial strain-induced displacements of Co and Pt atoms can alter the net DMI vector nearly twofold.[46] Dislocations produce atomic displacements around dislocation lines [light-blue regions in Figs. 4(c) and 4(d)], altering the DMI vector. Additionally, atomic displacements affect 3*d*-5*d* orbital hybridization near the Fermi energy, which correlates strongly with the DMI.[16] The asymmetric dislocation distribution within compositional gradient regions generates a net bulk DMI through these combined effects. While we could not determine dislocation-type-dependent contributions with sufficient statistical significance, different dislocation types likely affect the DMI differently. Interestingly, the asymmetric strain field distributions observed here may also relate to other phenomena: recent findings show that nominally symmetric bulk FeNi alloys exhibit substantial bulk spin-orbit torque induced by asymmetric strain distributions,[55] suggesting broader implications for spin-orbit physics.

We note that defect density is known to increase with sputtering pressure,[27,53] suggesting that an even more pronounced asymmetric dislocation distribution and consequently a stronger bulk DMI is to be expected in the higher-pressure samples, consistent with the enhanced Néel-type spin textures observed by LTEM. Furthermore, the fact that a strongly biased dislocation distribution is already clearly demonstrated in the 0.4-sample, which exhibits the weakest DMI in our series, provides a conservative yet compelling demonstration of the proposed defect-mediated mechanism.

Although we demonstrate that asymmetric defect distributions generate a net bulk DMI, several open questions remain. The specific contribution of each type and spatial distribution of defects to the net DMI, as well as the sign of the DMI at higher sputtering pressures, have yet to be determined. Furthermore, rigorous quantitative determination of the DMI magnitude — for example, through stripe domain periodicity analysis using established multilayer models, Brillouin light scattering spectroscopy, or domain wall dynamics measurements under in-plane magnetic fields — will be essential to establish a quantitative connection between the defect distribution and the resulting DMI. Resolving these issues will require controlled defect engineering combined with such quantitative DMI measurements and complementary theoretical modeling of defect-mediated spin–orbit interactions.

In conclusion, we have elucidated the microscopic origin of substantial DMI in nominally symmetric Pt/Co/Pt multilayers. By systematically varying the sputter gas pressure, we demonstrate enhanced net DMI that stabilizes chiral spin textures, despite the ill-defined interfaces. We reveal that macroscopic compositional asymmetry and the SOC enhancement are insufficient to account for the observed DMI. Instead, we identify a strongly biased distribution of dislocations within compositional gradient regions, which breaks inversion symmetry and consequently generates the bulk DMI through defect-induced atomic displacements within the three-site Fert–Levy framework. This defect-mediated mechanism resolves the long-standing puzzle of sizable DMI in nominally symmetric multilayers and provides that defect distributions may offer a pathway for tuning DMI in sputtered thin films.

**ACKNOWLEDGMENT**

We thank K. Higuchi, Y. Yamamoto, N. Ikarashi, and H. Cheong for technical support and discussion. This work was supported by JSPS KAKENHI Grants (No. 21H01029, No. 25K21662, and No. 25K07820) and JST Adaptable and Seamless Technology transfer Program through Target-driven R&D (A-STEP) (No. JPMJTR24R7). A part of this work was also supported by "Advanced Research Infrastructure for Materials and Nanotechnology in Japan (ARIM)" of the Ministry of Education, Culture, Sports, Science and Technology (MEXT) of Japan.

**REFERENCES**

1. S. Emori, U. Bauer, S. M. Ahn, E. Martinez, and G. S. Beach, “Current-driven dynamics of chiral ferromagnetic domain walls,” Nat. Mater. **12**, 611–616 (2013).
2. K. S. Ryu, L. Thomas, S. H. Yang, and S. Parkin, “Chiral spin torque at magnetic domain walls,” Nat. Nanotechnol. **8**, 527–533 (2013).
3. W. Jiang, P. Upadhyaya, W. Zhang, G. Yu, M. B. Jungfleisch, F. Y. Fradin, J. E. Pearson, Y. Tserkovnyak, K. L. Wang, O. Heinonen, S. G. E. te Velthuis, and A. Hoffmann, “Blowing magnetic skyrmion bubbles,” Science **349**, 283–286 (2015).
4. S. Woo, K. Litzius, B. Krüger, M. Im, L. Caretta, K. Richter, M. Mann, A. Krone, R. Reeve, M. Weigand, P. Agrawal, I. Lemesh, M. Mawass, Peter. Fischer, M. Kläui, and G. S. D. Beach, “Observation of room-temperature magnetic skyrmions and their current-driven dynamics in ultrathin metallic ferromagnets,” Nat. Mater. **15**, 501–506 (2016).
5. C. Moreau-Luchaire, C. Moutafis, N. Reyren, J. Sampaio, C. A. F. Vaz, N. Van Horne, K. Bouzehouane, K. Garcia, C. Deranlot, P. Warnicke, P. Wohlhüter, J.-M. George, M. Weigand, J. Raabe, V. Cros, and A. Fert, “Additive interfacial chiral interaction in multilayers for stabilization of small individual skyrmions at room temperature,” Nat. Nanotechnol. **11**, 444–448 (2013).
6. O. Boulle, J. Vogel, H. Yang, S. Pizzini, D. de Souza Chaves, A. Locatelli, T. O. Menteş, A. Sala, L. D. Buda-Prejbeanu, O. Klein, M. Belmeguenai, Y. Roussigné, A. Stashkevich, S. M. Chérif, L. Aballe, M. Foerster, M. Chshiev, S. Auffret, I. M. Miron, and G. Gaudin, “Room-temperature chiral magnetic skyrmions in ultrathin magnetic nanostructures,” Nat. Nanotechnol. **11**, 449–454 (2016).
7. G. Chen, T. Ma, A. T. N’Diaye, H. Kwon, C. Won, Y. Wu, and A. K. Schmid, “Tailoring the chirality of magnetic domain walls by interface engineering,” Nat. Commun. **4**, 2671 (2013).

8. O. Boulle, S. Rohart, L.D. Buda-Prejbeanu, E. Jue, I. M. Miron, S. Pizzini, J. Vogel, G. Gaudin, and A. Thiaville, "Domain Wall tilting in the presence of the Dzyaloshinskii-Moriya interaction in out-of-plane magnetized magnetic nanotracks," Phys. Rev. Lett. **111**, 217203 (2013).
9. G. Yu, P. Upadhyaya, X. Li, W. Li, S. K. Kim, Y. Fan, K. L. Wong, Y. Tserkovnyak, P. K. Amiri, and K. L. Wang, "Room-temperature creation and spin–orbit torque manipulation of skyrmions in thin films with engineered asymmetry," Nano Lett. **16**, 1981–1988 (2016).
10. H. Imamura, T. Nozaki, S. Yuasa, and Y. Suzuki, "Deterministic magnetization switching by voltage control of magnetic anisotropy and Dzyaloshinskii-Moriya interaction under an in-plane magnetic field," Phys. Rev. Appl. **10**, 054039 (2018).
11. H. Wu, J. Nance, S. A. Razavi, D. Lujan, B. Dai, Y. Liu, H. He, B. Cui, D. Wu, K. Wong, K. Sobotkiewich, X. Li, G. P. Carman, and K. L. Wang, "Chiral symmetry breaking for deterministic switching of perpendicular magnetization by spin-orbit torque," Nano Lett. **21**, 515–521 (2021).
12. H. Wang, J. Chen, T. Liu, J. Zhang, K. Baumgaertl, C. Guo, Y. Li, C. Liu, P. Che, S. Tu, S. Liu, P. Gao, X. Han, D. Yu, M. Wu, D. Grundler, and H. Yu, "Chiral spin-wave velocities induced by all-garnet interfacial Dzyaloshinskii-Moriya interaction in ultrathin yttrium iron garnet films," Phys. Rev. Lett. **124**, 027203 (2020).
13. M. Küß, M. Heigl, L. Flacke, A. Hörner, M. Weiler, M. Albrecht, and A. Wixforth, "Nonreciprocal Dzyaloshinskii–Moriya magnetoacoustic waves," Phys. Rev. Lett. **125**, 217203 (2020).
14. K-W. Kim, H-W. Lee, K-J. Lee, and M. D. Stiles, "Chirality from interfacial spin-orbit coupling effects in magnetic bilayers," Phys. Rev. Lett. **111**, 216601 (2013)
15. H. X. Yang, A. Thiaville, S. Rohart, A. Fert, and M. Chshiev, "Anatomy of Dzyaloshinskii-Moriya interaction at Co/Pt interfaces," Phys. Rev. Lett. **115**, 267210 (2015).
16. A. Belabbes, G. Bihlmayer, F. Bechstedt, S. Blügel, and A. Manchon, "Hund's rule-driven Dzyaloshinskii-Moriya interaction at $3d$−$5d$ Interfaces," Phys. Rev. Lett. **117**, 247202 (2016).
17. M. Hoffmann, B. Zimmermann, G. P. Müller, D. Schürhoff, N. S. Kiselev, C. Melcher, and S. Blügel, "Antiskyrmions stabilized at interfaces by anisotropic Dzyaloshinskii-Moriya interactions," Nat. Commun. **8**, 308 (2017).
18. H. Yang, G. Chen, A. A. C. Cotta, A. T. N'Diaye, S. A. Nikolaev, E. A. Soares, W. A. A. Macedo, K. Liu, A. K. Schmid, A. Fert, and M. Chshiev, "Significant Dzyaloshinskii-Moriya interaction at graphene-ferromagnet interfaces due to the Rashba effect," Nat. Mater. **17**, 605–609 (2018).
19. B. Yang, Q. Cui, J. Liang, M. Chshiev, and H. Yang, "Reversible control of Dzyaloshinskii-Moriya interaction at the graphene/Co interface via hydrogen absorption," Phys. Rev. B **101**, 014406 (2020).
20. C. Deger, "Strain-enhanced Dzyaloshinskii–Moriya interaction at Co/Pt interfaces," Sci. Rep. **10**, 12314 (2020).
21. H. Yang, J. Liang, and Q. Cui, "First-principles calculations for Dzyaloshinskii–Moriya interaction," Nat. Rev. Phys. **5**, 43–61 (2023).
22. A. Hrabec, N. A. Porter, A. Wells, M. J. Benitez, G. Burnell, S. McVitie, D. McGrouther, T. A. Moore, and C. H. Marrows, "Measuring and tailoring the Dzyaloshinskii-Moriya interaction in perpendicularly magnetized thin films," Phys. Rev. B **90**, 020402(R) (2014).
23. S.-G. Je, D.-H. Kim, S.-C. Yoo, B.-C. Min, K.-J. Lee, and S.-B. Choe, "Asymmetric magnetic domain-wall motion by the Dzyaloshinskii-Moriya interaction," Phys. Rev. B **88**, 214401 (2013).
24. S. D. Pollard, J. A. Garlow, J. Yu, Z. Wang, Y. Zhu, and H. Yang, "Observation of stable Néel skyrmions in

cobalt/palladium multilayers with Lorentz transmission electron microscopy," Nat. Commun. **8**, 14761 (2017).

25. J. Brandão, D. A. Dugato, R. L. Seeger, J. C. Denardin, T. J. A. Mori, and J. C. Cezar, "Observation of magnetic skyrmions in unpatterned symmetric multilayers at room temperature and zero magnetic field," Sci. Rep. **9**, 4144 (2019).
26. Y. Wei, C. Liu, Z. Zeng, X. Wang, J. Wang, and Q. Liu, "Room-temperature zero field and high-density skyrmions in Pd/Co/Pd multilayer films," J. Magn. Magn. Mater. **521**, 167507 (2021).
27. R. Lavrijsen, D. M. F. Hartmann, A. van den Brink, Y. Yin, B. Barcones, R. A. Duine, M. A. Verheijen, H. J. M. Swagten, and B. Koopmans, "Asymmetric magnetic bubble expansion under in-plane field in Pt/Co/Pt: Effect of interface engineering," Phys. Rev. B **91**, 104414 (2015).
28. A. W. J. Wells, P. M. Shepley, C. H. Marrows, and T. A. Moore, "Effect of interfacial intermixing on the Dzyaloshinskii-Moriya interaction in Pt/Co/Pt," Phys. Rev. B **95**, 054428 (2017).
29. A. L. Balk, K-W. Kim, D. T. Pierce, M. D. Stiles, J. Unguris, and S. M. Stavis, "Simultaneous control of the Dzyaloshinskii-Moriya interaction and magnetic anisotropy in nanomagnetic trilayers," Phys. Rev. Lett. **119**, 077205 (2017).
30. B. Zimmermann, W. Legrand, D. Maccariello, N. Reyren, V. Cros, S. Blügel, and A. Fert, "Dzyaloshinskii-Moriya interaction at disordered interfaces from ab initio theory: Robustness against intermixing and tunability through dusting," Appl. Phys. Lett. **113**, 232403 (2018).
31. P. C. Carvalho, I. P. Miranda, J. Brandão, A. Bergman, J. C. Cezar, A. B. Klautau, and H. M. Petrilli, "Correlation of interface interdiffusion and skyrmionic phases," Nano Lett. **23**, 4854–4861 (2023).
32. D-H. Kim, M. Haruta, H-W. Ko, G. Go, H-J. Park, T. Nishimura, D-Y. Kim, T. Okuno, Y. Hirata, Y. Futakawa, H. Yoshikawa, W. Ham, S. Kim, H. Kurata, A. Tsukamoto, Y. Shiota, T. Moriyama, S-B. Choe, K-J. Lee, and T. Ono, "Bulk Dzyaloshinskii–Moriya interaction in amorphous ferrimagnetic alloys," Nat. Mater. **18**, 685–690 (2019).
33. S. Krishnia, E. Haltz, L. Berges, L. Aballe, M. Foerster, L. Bocher, R. Weil, A. Thiaville, J. Sampaio, and A. Mougin, "Spin-orbit coupling in single-layer ferrimagnets: Direct observation of spin-orbit torques and chiral spin textures," Phys. Rev. Applied **16**, 024040 (2021).
34. J. Liang, M. Chshiev, A. Fert, and H. Yang, "Gradient-induced Dzyaloshinskii−Moriya interaction," Nano Lett. **22**, 10128–10133 (2022).
35. Q. Zhang, J. Liang, K. Bi, L. Zhao, H. Bai, Q. Cui, H. A. Zhou, H. Bai, H. Feng, W. Song, G. Chai, O. Gladii, H. Schultheiss, T. Zhu, J. Zhang, Y. Peng, H. Yang, and W. Jiang, "Quantifying the Dzyaloshinskii-Moriya interaction induced by the bulk magnetic asymmetry," Phys. Rev. Lett. **128**, 167202 (2022).
36. J. Park, T. Kim, G. W. Kim, V. Bessonov, A. Telegin, I. G. Iliushin, A. A. Pervishko, D. Yudin, A. Y. Samardak, A. V. Ognev, A. S. Samardak, J. Cho, and Y. K. Kim, "Compositional gradient induced enhancement of Dzyaloshinskii–Moriya interaction in Pt/Co/Ta heterostructures modulated by Pt–Co alloy intralayers," Acta Mater. **241**, 118383 (2022).
37. A. Michels, D. Mettus, I. Titov, A. Malyeyev, M. Bersweiler, P. Bender, I. Peral, R. Birringer, Y. Quan, P. Hautle, and J. Kohlbrecher, "Microstructural-defect-induced Dzyaloshinskii-Moriya interaction," Phys. Rev. B **99**, 014416 (2019).
38. A. Chakraborty, A. K. Srivastava, A. K. Sharma, A. K. Gopi, K. Mohseni, A. Ernst, H. Deniz, B. K. Hazra, S. Das, P. Sessi, and I. Kostanovskiy, "Magnetic skyrmions in a thickness tunable 2D ferromagnet from a defect driven Dzyaloshinskii-Moriya interaction," Adv. Mater. **34**, 2108637 (2022).

39. Z. Li, H. Zhang, G. Li, J. Guo, Q. Wang, Y. Deng, Y. Hu, X. Hu, C. Liu, M. Qin, X. Shen, R. Yu, X. Gao, Z. Liao, J. Liu, Z. Hou, Y. Zhu, and X. Fu, "Room-temperature sub-100 nm Néel-type skyrmions in non-stoichiometric van der Waals ferromagnet $Fe_{3-x}GaTe_2$ with ultrafast laser writability," Nat. Commun. **15**, 1017 (2024).
40. C. Zhang, Z. Jiang, J. Jiang, W. He, J. Zhang, F. Hu, S. Zhao, D. Yang, Y. Liu, Y. Peng, H. Yang, and H. Yang, "Above-room-temperature chiral skyrmion lattice and Dzyaloshinskii–Moriya interaction in a van der Waals ferromagnet $Fe_{3-x}GaTe_2$," Nat. Commun. **15**, 4472 (2024).
41. A. Fert and P. M. Levy, "Role of anisotropic exchange interactions in determining the properties of spin-glasses," Phys. Rev. Lett. **44**, 1538 (1980).
42. A. Thiaville, S. Rohart, É. Jué, V. Cros, and A. Fert, "Dynamics of Dzyaloshinskii domain walls in ultrathin magnetic films," EPL **100**, 57002 (2012).
43. T. Karino, D. Shimizu, A. P. Ohki, N. Ikarashi, T. Kato, D. Oshima, and M. Nagao, "Domain-stored skyrmion structures for a reading error-detectable racetrack memory," arXiv:2407.08527
44. W. Legrand, J. Y. Chauleau, D. Maccariello, N. Reyren, S. Collin, K. Bouzehouane, N. Jaouen, V. Cros, and A. Fert, "Hybrid chiral domain walls and skyrmions in magnetic multilayers," Sci. Adv. **4**, eaat0415 (2018).
45. A. Vansteenkiste, J. Leliaert, M. Dvornik, M. Helsen, F. Garcia-Sanchez, B. Van Waeyenberge, "The design and verification of MuMax3," AIP Adv. **4**, 107133 (2014).
46. N. S. Gusev, A. V. Sadovnikov, S. A. Nikitov, M. V. Sapozhnikov, and O. G. Udalov, "Manipulation of the Dzyaloshinskii–Moriya interaction in Co/Pt multilayers with strain," Phys. Rev. Lett. **124**, 157202 (2020).
47. I.-H. Kao, R. Muzzio, H. Zhang, M. Zhu, J. Gobbo, S. Yuan, D. Weber, R. Rao, J. Li, J. H. Edgar, J. E. Goldberger, J. Yan, D. G. Mandrus, J. Hwang, R. Cheng, J. Katoch, and S. Singh, "Deterministic switching of a perpendicularly polarized magnet using unconventional spin–orbit torques in $WTe_2$," Nat. Mater. **21**, 1029–1034 (2022).
48. J. W. Lee, Y.-W. Oh, S.-Y. Park, A. I. Figueroa, G. van der Laan, G. Go, K.-J. Lee, and B.-G. Park, "Enhanced spin-orbit torque by engineering Pt resistivity in Pt/Co/$AlO_x$ structures," Phys. Rev. B **96**, 064405 (2017).
49. R. Wang, Z. Xiao, H. Liu, Z. Quan, X. Zhang, M. Wang, M. Wu, and X. Xu, "Enhancement of perpendicular magnetic anisotropy and spin-orbit torque in Ta/Pt/Co/Ta multi-layered heterostructures through interfacial diffusion," Appl. Phys. Lett. **114**, 042404 (2019).
50. D. K. Ojha, R. Chatterjee, Y.-L. Lin, Y.-H. Wu, P.-W. Chen, and Y.-C. Tseng, "Spin-torque efficiency enhanced in sputtered topological insulator by interface engineering," J. Magn. Magn. Mater. **572**, 170638 (2023).
51. L. Chen, K. Zhang, B. Li, B. Hong, W. Huang, Y. He, X. Feng, Z. Zhang, K. Lin, W. Zhao, and Y. Zhang, "Engineering symmetry breaking enables efficient bulk spin-orbit torque-driven perpendicular magnetization switching," Adv. Funct. Mater. **34**, 2308823 (2023).
52. B. T. Thole and G. van der Laan, "Branching ratio in x-ray absorption spectroscopy," Phys. Rev. B **38**, 3158 (1988).
53. R. Messier, A. P. Giri, and R. A. Roy, "Revised structure zone model for thin film physical structure," J. Vac. Sci. Technol. A **2**, 500–503 (1984).
54. S. Yuasa, T. Nagahama, A. Fukushima, Y. Suzuki, and K. Ando, "Giant room-temperature magnetoresistance in single-crystal Fe/MgO/Fe magnetic tunnel junctions," Nat. Mater. **3**, 868–871 (2004).
55. R. E. Maizel, S. Wu, P. P. Balakrishnan, A. J. Grutter, C. J. Kinane, A. J. Caruana, P. Nakarmi, B. Nepal, D. A. Smith, Y. Lim, J. L. Jones, W. C. Thomas, J. Zhao, F. M. Michel, T. Mewes, and S. Emori, "Vertically graded Fe-Ni alloys with low damping and a sizable spin-orbit torque," Phys. Rev. Applied **22**, 044052 (2024).

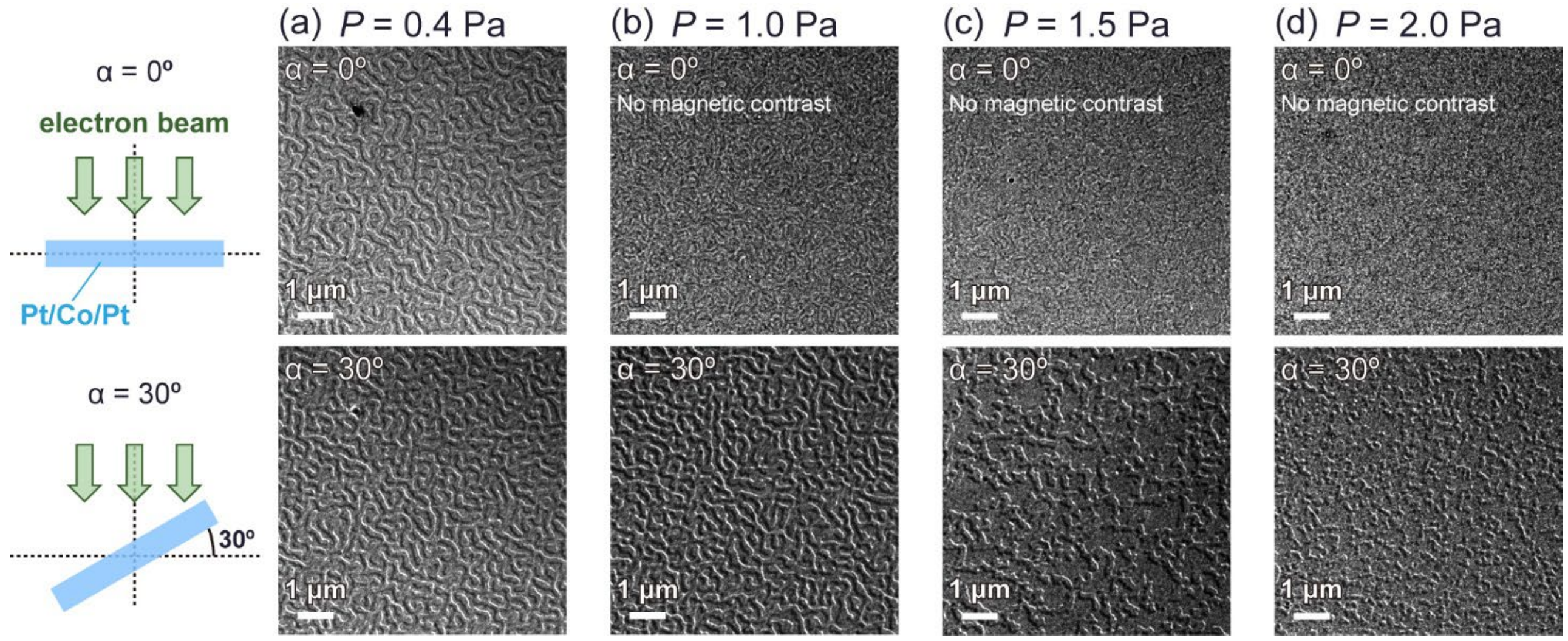

**FIG. 1.** LTEM images in the samples deposited with $P$ = 0.4 (a), 1.0 (b), 1.5 (c), and 2.0 Pa (d). The top and bottom row images were captured at sample tilt angles $\alpha$ = 0° and 30°, respectively, as shown in the left-side schematics. In (b)–(d), the line and circle-like contrasts only at $\alpha$ = 30° represent Néel-type DWs and skyrmions, respectively.

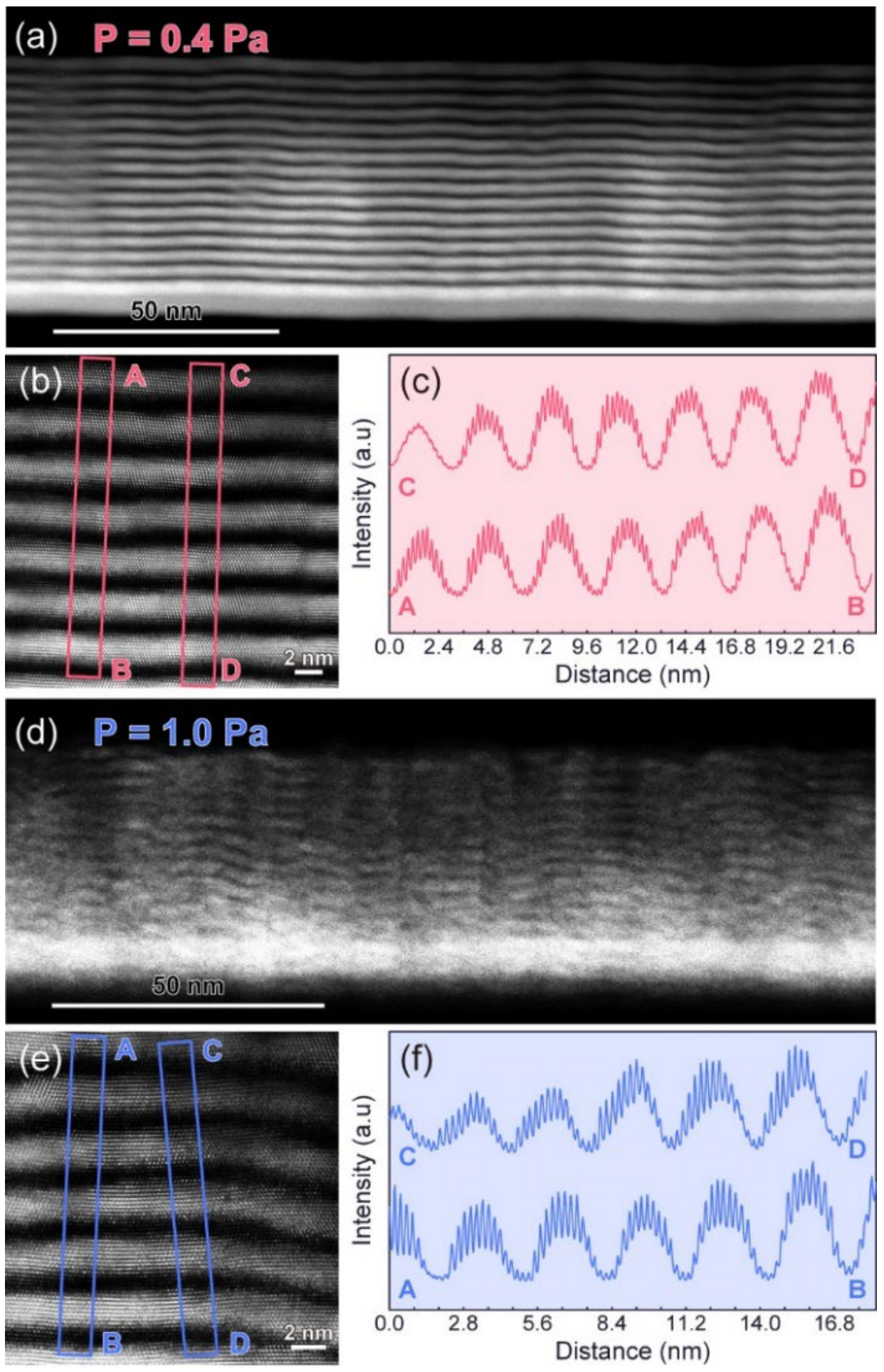

**FIG. 2.** Cross-sectional HAADF-STEM images in the 0.4-sample (a),(b) and the 1.0-sample (d),(e). (b),(e) High magnification images. (c),(f) Integrated intensity profiles obtained from the rectangular regions along A-B and C-D in (b) and (e). The sharp peaks correspond to atomic columns or lattice fringes.

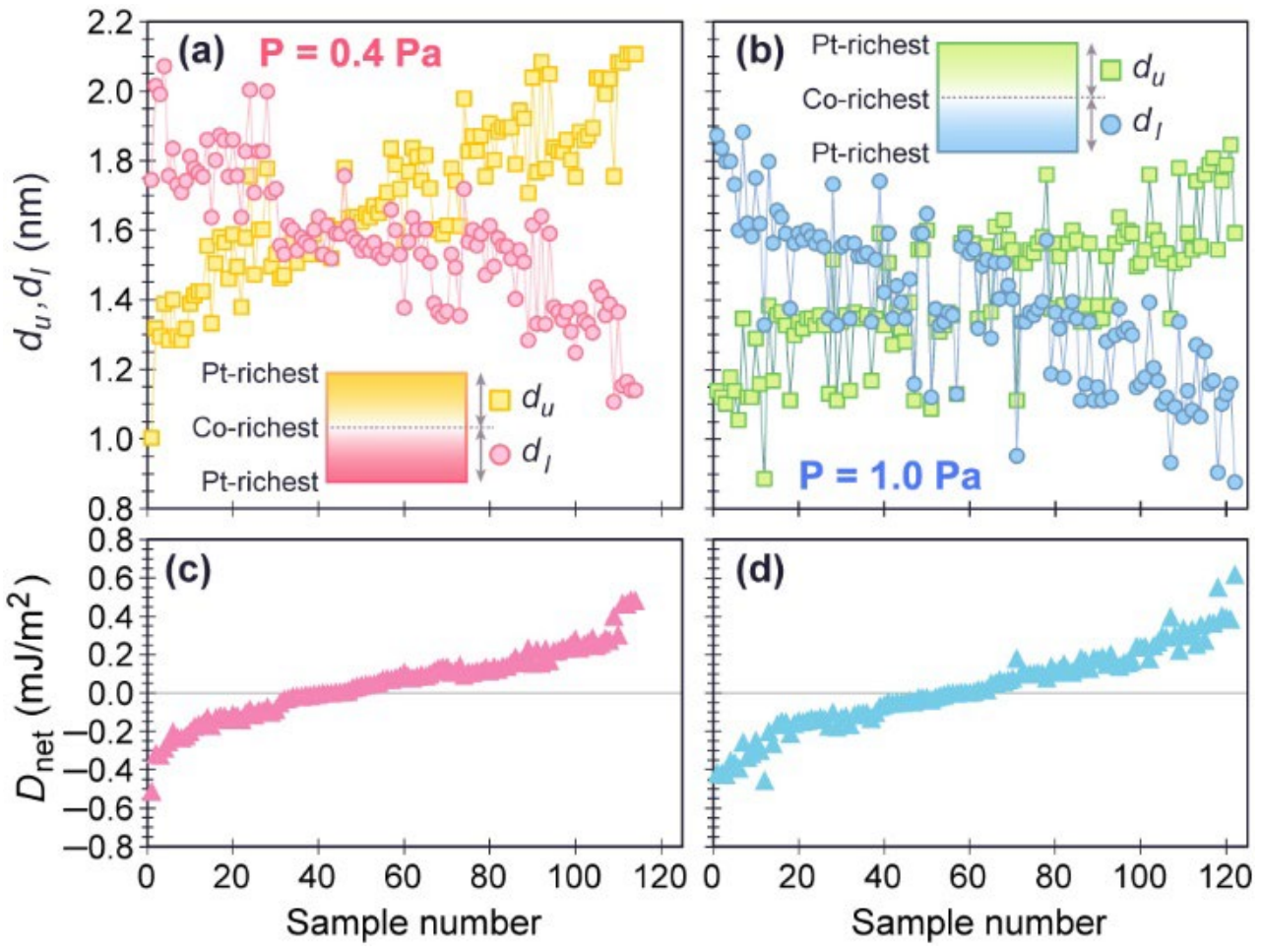

**FIG. 3.** Compositional gradient distances $d_u$ (squares) and $d_l$ (circles) for the individual regions. Intensity profiles were obtained using the line width $w$ = 1.88 nm in 114 and 122 regions for the 0.4-sample (a) and 1.0-sample (b), respectively. The insets depict the definitions of $d_u$ and $d_l$. The regions (sample number) are shown in descending order of their $d_l$ values. (c),(d) Net bulk DMI energy $D_{net}$ for the individual regions corresponding to (a) and (b), respectively. The estimation is based on a previous study.[35]

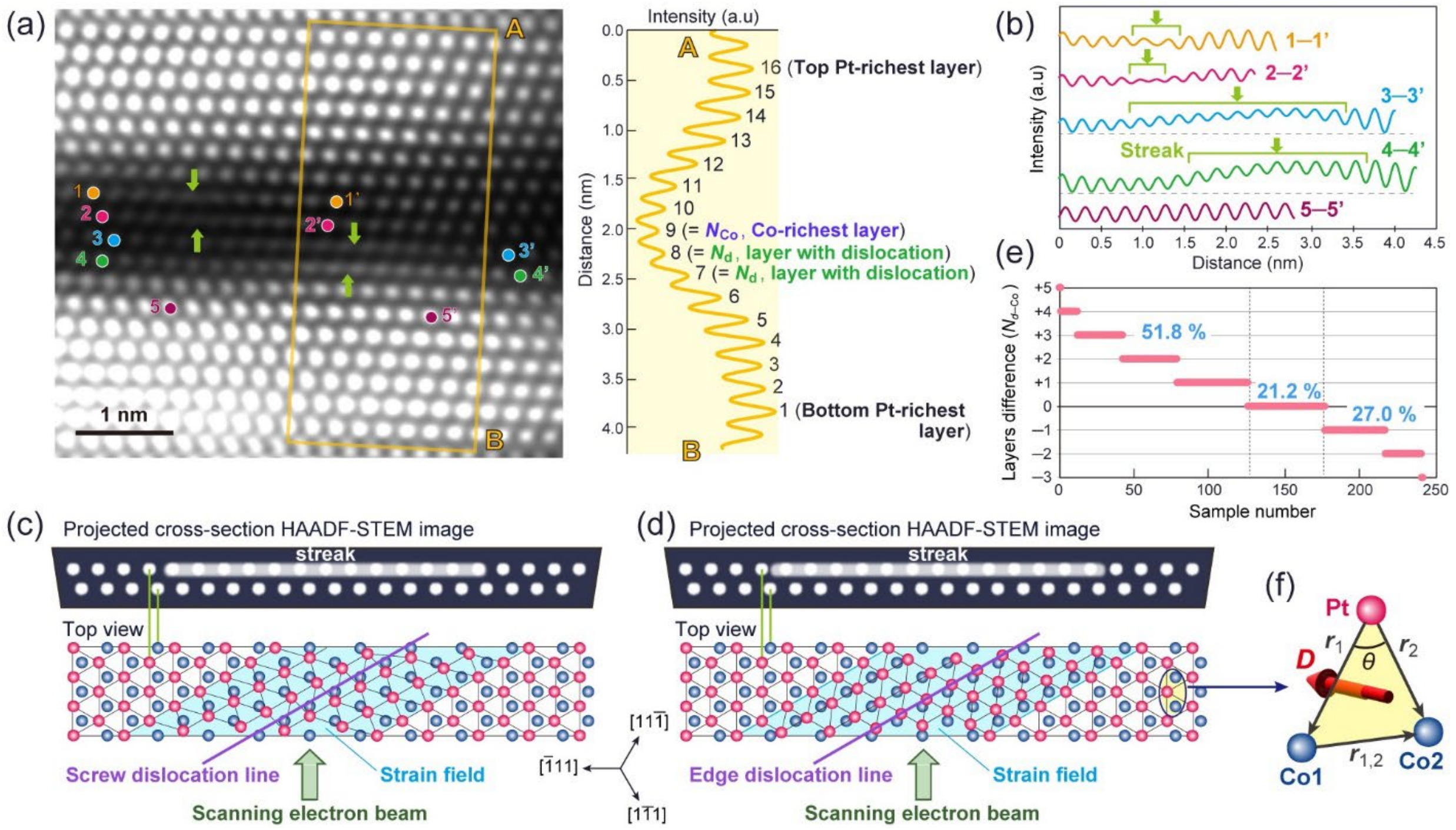

**FIG. 4.** Dislocation position analysis for the 0.4-sample. (a) Left: A representative Fourier-filtered cross-section HAADF-STEM image. The green arrows point to streak-like blurred contrasts. Right: Intensity profile along the A-B line in the left image. (b) Intensity profiles along the 1-1' to 5-5' lines in (a). (c),(d) Schematics of screw and edge dislocations, respectively. The bottom panels depict the schematics of the top view of the stacking structures with screw and edge dislocations. The red and blue balls represent the relatively Pt-rich upper layer and the relatively Co-rich lower layer, respectively. The top panels depict the schematics of the projected cross-section HAADF-STEM images, corresponding to the bottom panels. (e) Analysis of the layer differences $N_{d\text{-}Co} = N_d - N_{Co}$. The dislocation distribution shows clear

asymmetry with 51.8% positive (above the Co-richest atomic layers), 26.6% negative (below), and 21.6% zero (coincidence) values. (f) Schematic of the three-site Fert-Levy model.[41] $\boldsymbol{D}$ is the DMI vector.

## Supplementary Material

**CONTENTS**

## I. Considerable intermixing near the grain boundaries and the composition bending regions

We observe considerable intermixing near the grain boundaries and the composition bending region in the 0.4- and 1.0-samples. Figures S1(a)-(d) and S1(e)-(h) display HAADF-STEM images of the 0.4- and 1.0-samples, respectively. The yellow arrows indicate the location of the grain boundaries, and the blue squares enclose the regions of the compositional bending. Considerable intermixing occurs near the grain boundaries in both the 0.4- and 1.0-samples. In the 1.0-sample, the compositional bending near the grain boundary is particularly noticeable. (The lattice planes are relatively flat.) On the other hand, as depicted in Fig. S1(d), intermixing also occurs near the grain boundaries where the compositional bending is less, suggesting that intermixing near the grain boundaries is inherent regardless of the compositional bending. The extent of intermixing decreases away from the grain boundaries. The compositional bending regions are observed within the grains, leading to intermixing.

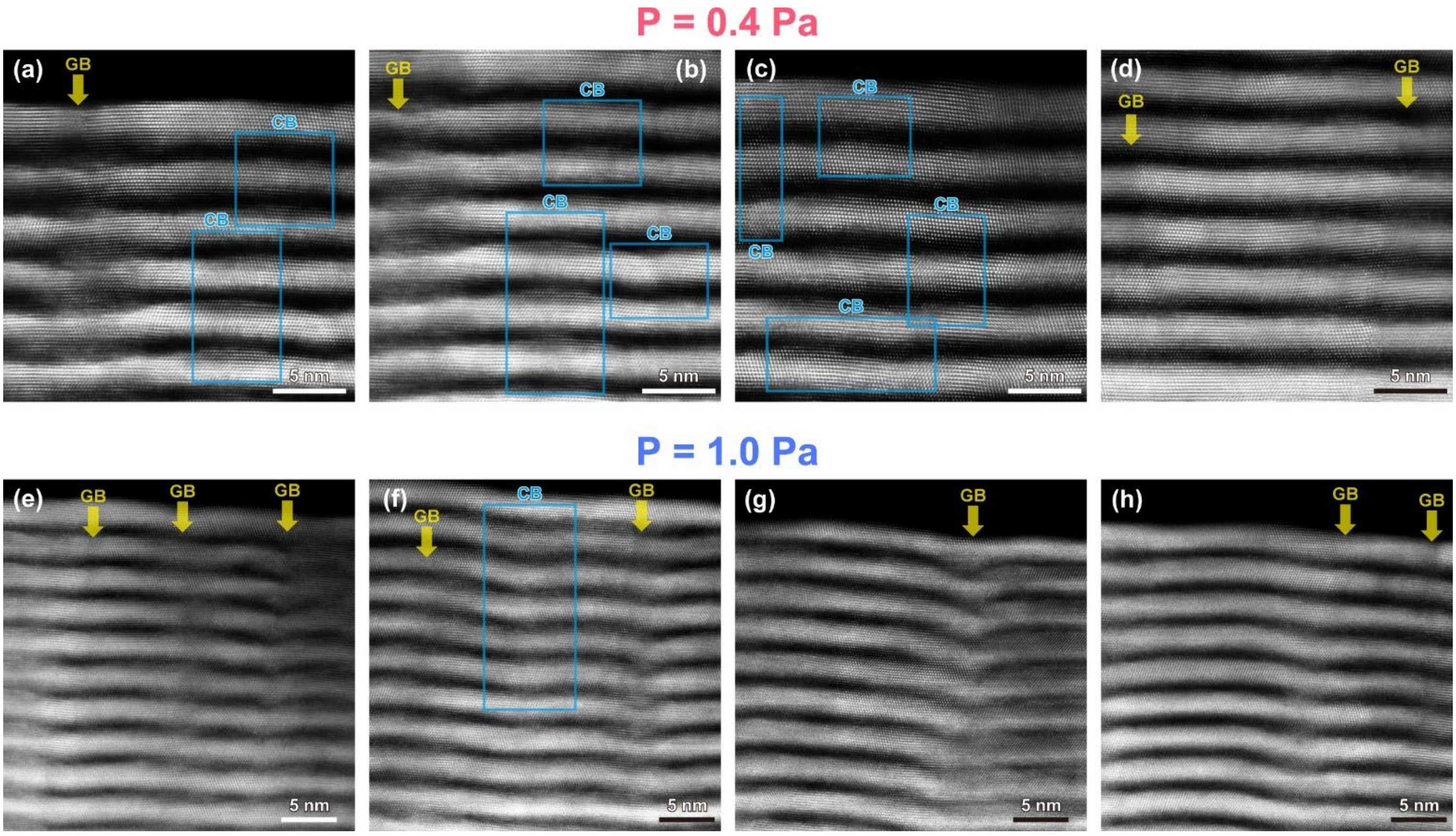

FIG. S1. HAADF-STEM images in in the 0.4-sample (a)-(d) and the 1.0-sample (e)-(h). The yellow arrows indicate the location of the grain boundaries (GBs), and the blue squares enclose the regions of the compositional bending (CB).

We also note that, although interface roughness and local curvature of the composition bending regions could in principle generate local DMI contributions through mechanisms analogous to those discussed for geometrically asymmetric interfaces, such contributions would only survive macroscopic averaging if they were statistically biased in a preferred direction. As is evident from Figs. S1(a)–S1(h), the composition bending exhibits both upward-dome and downward-dome configurations with no systematic directional preference. Consequently, any local DMI contribution from upward-curved regions would be compensated by an equal and opposite contribution from downward-curved regions, yielding a negligible net DMI when averaged over the sample area. Interface steps distributed without a preferred orientation would similarly cancel. This is in marked contrast to the asymmetric dislocation distribution demonstrated in Fig. 4(e) of the main text, which exhibits a clear and statistically significant directional bias (51.8% above versus 26.6% below the Co-richest atomic layers) that survives macroscopic averaging and is therefore capable of generating a net bulk DMI. We thus conclude that statistically unbiased interface roughness and composition bending are unlikely to be the dominant source of the observed net DMI, and that the asymmetric dislocation distribution constitutes the primary symmetry-breaking mechanism.

## II. Evaluation of the critical DMI constant and domain wall chirality profile in the thick-film limit

As discussed in the main text, stabilizing a purely Néel-type domain wall requires a sufficiently large DMI to overcome the magnetostatic shape anisotropy of the DW cross-section. To quantify this threshold, we use the analytical model of Thiaville et al.[S1]:

$$D_{\mathrm{c}} = \frac{4}{\pi}\sqrt{\frac{A}{K_{\mathrm{u}}}} \cdot \left[N_x\left(\frac{1}{2}\mu_0 M_{\mathrm{s}}^2\right)\right],$$

where $A$ is the exchange stiffness, $K_{\mathrm{u}}$ is the uniaxial magnetic anisotropy, $M_{\mathrm{s}}$ is the saturation magnetization, and $N_x$ is the transverse demagnetizing factor of the DW cross-section. The factor $N_x$ quantifies the shape anisotropy penalty for the Néel (in-plane transverse) versus Bloch (in-plane axial) magnetization configuration within the wall.

The value of $N_x$ is a sensitive function of the ratio of film thickness t to DW width $\pi\Delta$ (where $\Delta = \sqrt{A/K_{\mathrm{u}}}$). In the ultra-thin film limit ($t \ll \pi\Delta$), as in the Co ($t = 0.6$ nm) system of Thiaville et al., the DW cross-section resembles a thin ribbon and $N_x$ is small ($\approx 0.0224$). This value can be understood from the analytical relation $N_x \approx t \cdot \ln 2/(\pi\Delta)$ valid in the limit $t \ll \pi\Delta$. In contrast, our multilayer samples have a total magnetic thickness of $t \approx 45$ nm, which exceeds the estimated DW width $\pi\Delta \approx 2.0$–$6.8$ nm by more than an order of magnitude. In this regime ($t \gg \pi\Delta$), a standard analytical approximation for a rectangular DW cross-section gives:

$$N_x \approx \frac{t}{(t + \pi\Delta)},$$

yielding $N_x \approx 0.87$–$0.96$ for our samples. We note that this is an analytical estimate based on a simplified rectangular cross-section model. The precise value of $N_x$ in a realistic multilayer DW may differ depending on the actual magnetization profile and interlayer coupling. However, the key result—that $N_x$ is considerably larger than 0.0224 for $t \approx 45$ nm—is not sensitive to these details, as the ratio $t/\pi\Delta \gg 1$ throughout the relevant parameter range.

Table S1 presents the resulting $D_{\mathrm{c}}$ values for all three samples and three values of $A$. The large $N_x$ in our thick-film geometry substantially increases the magnetostatic shape anisotropy penalty for Néel-type walls, requiring correspondingly large DMI values (approximately 1.0–3.5 mJ/m$^2$, depending on $A$) to stabilize the Néel configuration. These values should be interpreted as theoretical upper bounds under the assumption of a *purely Néel-type spin texture*; the actual DMI constants can be lower if the DW adopts an asymmetric Bloch/Néel hybrid structure, as mentioned in the main text.

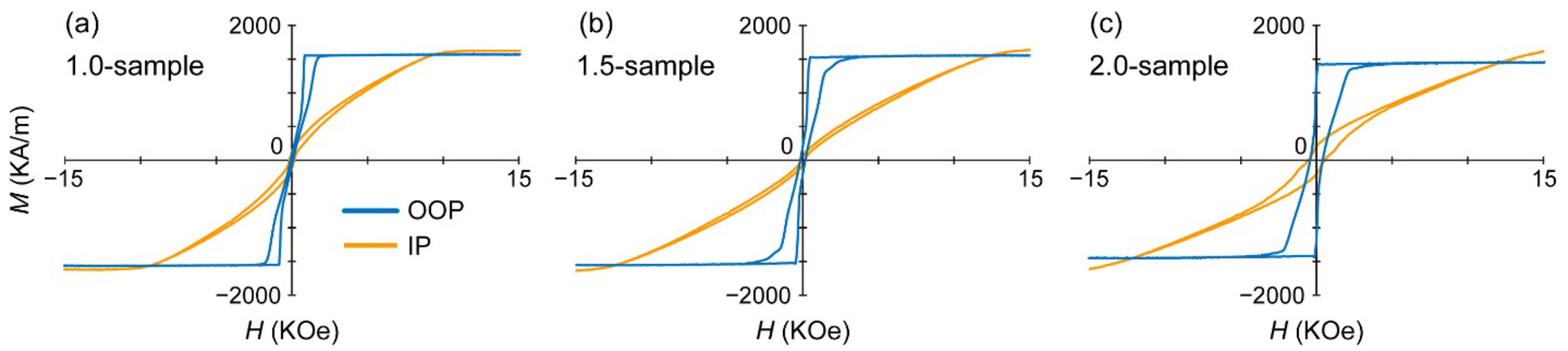

FIG. S2. Out-of-plane (OOP, blue) and in-plane (IP, orange) magnetization curves of 1.0- (a), 1.5- (b), and 2.0-samples (c).

Table S1. Calculated critical DMI constants ($D_c$) in the thick-film limit ($t \approx 45$ nm), using the thickness-corrected $N_x \approx t/(t + \pi\Delta)$.

| Sample | $M_s$ (KA/m) | $K_u$ (MJ/m$^2$) | $A$ (pJ/m) | $W = \pi\Delta$ (nm) | $N_x$ | $D_c$ (mJ/m$^2$) |
|---|---|---|---|---|---|---|
| 1.0-sample | 1520 | 2.12 | 1.0 | 2.2 | 0.954 | 1.21 |
| | | | 5.0 | 4.8 | 0.903 | 2.56 |
| | | | 10.0 | 6.8 | 0.868 | 3.48 |
| 1.5-sample | 1480 | 2.19 | 1.0 | 2.1 | 0.955 | 1.13 |
| | | | 5.0 | 4.7 | 0.905 | 2.39 |
| | | | 10.0 | 6.7 | 0.870 | 3.26 |
| 2.0-sample | 1430 | 2.51 | 1.0 | 2.0 | 0.958 | 0.99 |
| | | | 5.0 | 4.4 | 0.910 | 2.10 |
| | | | 10.0 | 6.3 | 0.878 | 2.87 |

To assess the actual spin texture of the DWs in our multilayers and its relationship to the LTEM observations, we performed micromagnetic simulations[S2] of the DW profile following the full-geometry methodology of Legrand et al.,[S3] in which all magnetic Co layers and nonmagnetic Pt spacers are explicitly included. The azimuthal angle $\varphi$ of the DW magnetization was evaluated at the center of each Co layer as a function of its position $z$ along the multilayer thickness, for DMI values $D = 0.1$–$3.9$ mJ/m$^2$ and exchange stiffness values $A = 1.0$, 5.0, and 10.0 pJ/m, for both the 0.4- and 1.0-samples. The results are shown in Fig. S3. All panels show a characteristic off-center profile in which $\varphi$ increases continuously from near 0° at the substrate side to values approaching 180° near the surface, with the transition shifting toward larger $z$ (the surface side) as $D$ increases. This behavior is consistent with the hybrid chiral domain wall structure described by Legrand et al.[S3]

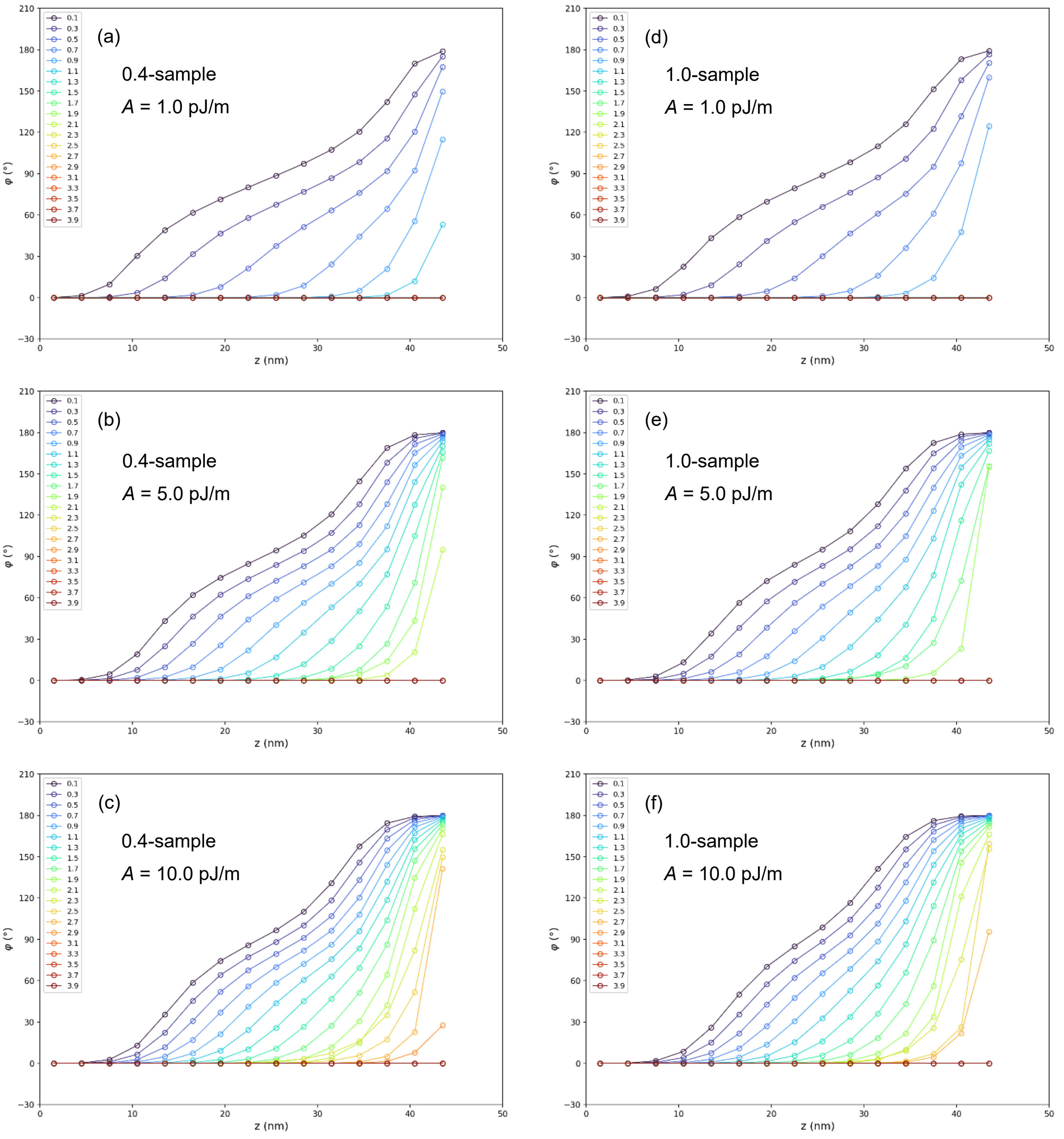

FIG. S3. Micromagnetic simulations of the azimuthal angle $\varphi$ of the DW magnetization as a function of layer position $z$ along the multilayer thickness, performed in the full geometry including all magnetic Co layers and nonmagnetic Pt spacers[S3], for the 0.4-sample (a)–(c) and the 1.0-sample (d)–(f), with $A$ = 1.0, 5.0, and 10.0 pJ/m, respectively. Each curve corresponds to a different $D$ (0.1–3.9 mJ/m$^2$), as indicated in the legend. $\varphi = 0°$ corresponds to a CCW Néel-type DW, $\varphi = 90°$ to a Bloch-type DW, and $\varphi = 180°$ to a CW Néel-type DW. The hybrid chiral DW structure, characterized by a continuous twist of $\varphi$ from $\approx 0°$ at the bottom to values approaching 180° near the top of the multilayer, is stabilized for the majority of the DMI range investigated.

Table S2. Thickness-averaged Bloch component $\langle \sin\varphi \rangle = (1/N)\sum \sin\varphi(z_i)$ calculated from the full-geometry micromagnetic simulations shown in Fig. S3, for the 0.4- and 1.0-samples and for $A$ = 1.0, 5.0, and 10.0 pJ/m. $N$ = 15 denotes the number of Co layers. $\langle \sin\varphi \rangle$ is proportional to the integrated electron phase shift responsible for LTEM contrast at $\alpha$ = 0°; values approaching unity indicate dominant Bloch character, while values approaching zero indicate suppression of the Bloch component within the hybrid chiral DW structure: $\langle \sin\varphi \rangle$ = 1 represents the formation of a pure Bloch-type DW, and $\langle \sin\varphi \rangle$ = 0 represents the formation of a pure Néel-type DW.

| 0.4-sample | | | |
|---|---|---|---|
| $D$ (mJ/m$^2$) | $A$ = 1.0 pJ/m | $A$ = 5.0 pJ/m | $A$ = 10.0 pJ/m |
| 0.1 | 0.5923 | 0.5051 | 0.4571 |
| 0.3 | 0.5214 | 0.4893 | 0.4491 |
| 0.5 | 0.3913 | 0.4588 | 0.4332 |
| 0.7 | 0.2476 | 0.4139 | 0.4096 |
| 0.9 | 0.1463 | 0.3574 | 0.3793 |
| 1.1 | 0.0694 | 0.2946 | 0.3426 |
| 1.3 | 0 | 0.2313 | 0.3007 |
| 2.1 | 0 | 0.0952 | 0.1452 |
| 2.3 | 0 | 0 | 0.166 |
| 3.1 | 0 | 0 | 0 |

| 1.0-sample | | | |
|---|---|---|---|
| $D$ (mJ/m$^2$) | $A$ = 1.0 pJ/m | $A$ = 5.0 pJ/m | $A$ = 10.0 pJ/m |
| 0.1 | 0.5546 | 0.4637 | 0.4119 |
| 0.3 | 0.4822 | 0.4479 | 0.4037 |
| 0.5 | 0.3549 | 0.4167 | 0.3876 |
| 0.7 | 0.2124 | 0.3715 | 0.3637 |
| 0.9 | 0.1253 | 0.3146 | 0.3325 |
| 1.1 | 0 | 0.2506 | 0.2954 |
| 1.3 | 0 | 0.1894 | 0.255 |
| 2.1 | 0 | 0.0001 | 0.1254 |
| 2.9 | 0 | 0 | 0 |
| 3.1 | 0 | 0 | 0 |

The simulations reveal that, for the majority of the DMI range investigated, the DW magnetization adopts a hybrid chiral configuration: the bottommost Co layers exhibit predominantly CCW Néel character ($\varphi \approx 0°$) defined by the DMI, intermediate layers show substantial Bloch character ($\varphi \approx 90°$), and the uppermost layers develop a CW Néel component ($90° < \varphi < 180°$), as a result of competition between the DMI and interlayer dipolar interactions. However, for sufficiently large $D$, the Bloch component is fully suppressed and the DW magnetization adopts a predominantly uniform Néel-type configuration throughout the multilayer thickness. The critical DMI value at which this transition occurs depends on both $A$ and the sample magnetic parameters, and is generally consistent with the $D_c$ values estimated in Table S1.

To connect these simulation results to the LTEM observations, we calculated the thickness-averaged Bloch component $\langle \sin\varphi \rangle = (1/N)\sum \sin\varphi(z_i)$, where $\varphi(z_i)$ is the azimuthal angle in the $i$-th Co layer and $N$ = 15. As noted in the main text, this quantity is proportional to the integrated electron phase shift that produces LTEM contrast at $\alpha$ = 0°. The results are summarized in Table S2. For the 0.4-sample with $D \approx$ 0.1–0.4 mJ/m$^2$ (reference values from Gusev et al.[S4] for unstrained Pt/Co/Pt), $\langle \sin\varphi \rangle$ is in the range 0.45–0.59, consistent with the observed Bloch-type LTEM contrast. As $D$ increases, $\langle \sin\varphi \rangle$ decreases for all values of $A$, reflecting the progressive suppression of the Bloch component. Once $\langle \sin\varphi \rangle$ falls below the detection threshold of our LTEM setup — whether due to the hybrid chiral spin texture at intermediate $D$ or to the approach of a uniform Néel-type configuration at large $D$ — the $\alpha$ = 0° contrast disappears. This DMI-driven decrease in $\langle \sin\varphi \rangle$ provides the physical mechanism underlying the observed transition from Bloch-type contrast in the 0.4-sample to its absence in the $P \geq 1.0$ Pa samples, as discussed in the main text.

### III. Quantitative evaluation of spin-orbit coupling by electron energy loss spectroscopy for the 0.4- and 1.0-samples

Electron energy loss spectroscopy (EELS) provides a direct measure of the effective spin-orbit coupling (SOC) of Co through the branching ratio $B$ at the Co $L_{2,3}$ edges. The core-loss spectra obtained by EELS are equivalent to those obtained by X-ray absorption spectroscopy (XAS), and recent XAS studies on Co- and Fe-based multilayers have demonstrated SOC enhancement through measurement of $B$, defined as the fraction of the total intensity $I$ at the $L_3$ edge: $B = I(L_3)/[I(L_3) + I(L_2)]$.[S5–S8] This branching ratio is theoretically defined as $B = B_0 + \langle l{\cdot}s/n_h \rangle$, where $B_0$ is the branching ratio without SOC, $l$ and $s$ are the orbital and spin angular momenta, respectively, and $n_h$ is the number of $d$-holes.[S9] An enhanced SOC leads to an increased branching ratio above the ideal statistical value of $B_0 \approx 0.667$, which corresponds to an $I(L_3)/I(L_2)$ ratio of 2:1.

We conducted EELS measurements on the 0.4- and 1.0-samples in TEM-EELS mode to obtain the Co $L_{2,3}$ core-loss spectra from the whole film regions, and additionally in STEM-EELS mode to obtain spectra from a grain boundary region in the 0.4-sample. The grain boundary of the 0.4-sample was selected for the STEM-EELS measurement because the larger grain size ($L \approx 20$ nm) of this sample allows reliable acquisition of spatially resolved spectra from the grain boundary region; the smaller grain size of the 1.0-sample ($L \approx 5$–$10$ nm) precludes such spatially selective measurements. The core-loss spectra are shown in Figs. S4(a)–S4(c), and the corresponding integration windows, integrated intensities $I(L_3)$ and $I(L_2)$, and branching ratios $B$ are summarized in Table S3.

The branching ratios obtained are $B = 0.66$ for the 0.4-sample (whole region), $B = 0.65$ for the 1.0-sample (whole region), and $B = 0.65$ for the grain boundary of the 0.4-sample. These values are close to the ideal statistical ratio of $B_0 \approx 0.667$ and are significantly lower than the enhanced branching ratios of $B = 0.75$–$0.82$ reported in previous studies of multilayer systems exhibiting substantial SOC enhancement.[S5,S6,S8] The near-identical branching ratios of the 0.4- and 1.0-samples — despite their markedly different DMI strengths — confirm that the effective SOC of Co is not significantly different between the two samples. Furthermore, the branching ratio at the grain boundary of the 0.4-sample is also consistent with those of the whole-film regions, indicating no significant SOC localization at grain boundaries. We therefore conclude that SOC enhancement cannot account for the substantially different DMI observed between the 0.4- and 1.0-samples.

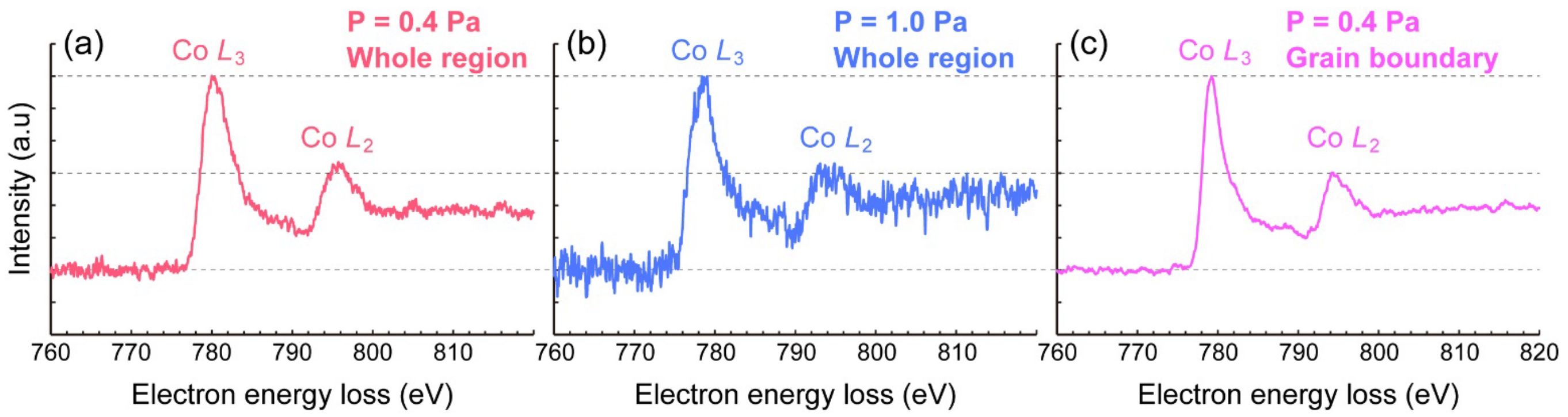

FIG. S4. Core-loss spectra in EELS obtained from the whole region of the 0.4-sample (a) and the 1.0-sample (b), and a grain boundary in the 0.4-sample (c). All spectra show nearly the same intensity and shape. The poor S/N ratio in (b) is probably because the thickness of the 1.0-sample is thicker than that of the 0.4-sample.

Table S3. Quantitative EELS analysis of the Co $L_{2,3}$ branching ratio $B$ for the 0.4- and 1.0-samples

| Sample | Measurement region | Integration window $L_3$ | Integrated intensity $I(L_3)$ | Integration window $L_2$ | Integrated intensity $I(L_2)$ | Branching ratio $B$ $(=I(L_3)/[I(L_3)+I(L_2)])$ |
|---|---|---|---|---|---|---|
| (a) 0.4-sample | Whole region | 777–791 eV | 378090.8046 | 792–800 eV | 194837.9732 | 0.6599 |
| (b) 1.0-sample | Whole region | 777–791 eV | 365829.1479 | 792–800 eV | 201039.8578 | 0.6454 |
| (c) 0.4-sample | Grain boundary | 777–791 eV | 320647.0196 | 792–800 eV | 175683.2107 | 0.6460 |

## IV. Numerical data of various distance measurements and number of layers for the 0.4- and 1.0-samples

Table S4. Numerical data of various distance measurements for the 0.4-sample. Deletion of duplicates is due to the presence of two or more dislocations in a single line profile. The sample numbers of Fig. 3(a) in the main text are rearranged in order of increasing $d_{l\text{-}u}$ value to make the dispersion easier to recognize.

| Sample number | Distance between the richest Pt at top and bottom | Remove duplicates from the left column | Distance of dislocations from the bottom richest Pt | Distance of the richest Co from the bottom richest Pt | Remove duplicates from the left column **[Fig. 3(a)]** | Distance of the richest Co from the top richest Pt **[Fig. 3(a)]** | Distance between $d_l$ and $d_u$ | Net Bulk DMI energy **[Fig. 3(c)]** | Distance between dislocations and the richest Co |
|---|---|---|---|---|---|---|---|---|---|
| | $d_{\text{Pt-Pt}}$ [nm] | $d_{\text{Pt-Pt}}$ [nm] | $d_d$ [nm] | $d_{\text{Co}}$ [nm] | $d_l$ [nm] | $d_u$ [nm] | $d_{l\text{-}u}$ [nm] | $D_{\text{net}}$ [mJ/m$^2$] | $d_{d\text{-Co}}$ [nm] |
| 1 | 2.993 | 2.993 | 2.145 | 1.709 | 1.709 | 1.284 | 0.425 | -0.23 | 0.436 |
| 2 | 2.993 | | 1.933 | 1.709 | | | | | 0.224 |
| 3 | 3.205 | 3.205 | 1.496 | 1.708 | 1.708 | 1.497 | 0.211 | -0.10 | -0.212 |
| 4 | 3.205 | | 1.296 | 1.708 | | | | | -0.412 |
| 5 | 3.181 | 3.181 | 2.557 | 1.708 | 1.708 | 1.473 | 0.235 | -0.11 | 0.849 |
| 6 | 3.181 | | 2.333 | 1.708 | | | | | 0.625 |
| 7 | 3.431 | 3.431 | 1.709 | 1.509 | 1.509 | 1.922 | -0.413 | 0.17 | 0.2 |
| 8 | 3.415 | 3.415 | 2.132 | 1.519 | 1.519 | 1.896 | -0.377 | 0.16 | 0.613 |
| 9 | 3.415 | | 1.92 | 1.519 | | | | | 0.401 |
| 10 | 3.415 | | 1.719 | 1.519 | | | | | 0.2 |
| 11 | 3.415 | | 1.519 | 1.519 | | | | | 0 |
| 12 | 3.415 | | 1.319 | 1.519 | | | | | -0.2 |
| 13 | 3.227 | 3.227 | 2.12 | 1.472 | 1.472 | 1.755 | -0.283 | 0.13 | 0.648 |
| 14 | 3.227 | | 1.908 | 1.472 | | | | | 0.436 |
| 15 | 3.227 | | 1.684 | 1.472 | | | | | 0.212 |
| 16 | 2.991 | 2.991 | 1.472 | 1.284 | 1.284 | 1.707 | -0.423 | 0.23 | 0.188 |
| 17 | 2.991 | | 1.284 | 1.284 | | | | | 0 |
| 18 | 2.991 | | 1.107 | 1.284 | | | | | -0.177 |
| 19 | 3.002 | 3.002 | 1.248 | 1.248 | 1.248 | 1.754 | -0.506 | 0.28 | 0 |
| 20 | 3.002 | | 1.048 | 1.248 | | | | | -0.2 |
| 21 | 3.229 | 3.229 | 2.581 | 1.508 | 1.508 | 1.721 | -0.213 | 0.10 | 1.073 |
| 22 | 3.229 | | 2.357 | 1.508 | | | | | 0.849 |
| 23 | 3.229 | | 1.993 | 1.508 | | | | | 0.485 |
| 24 | 3.229 | | 1.508 | 1.508 | | | | | 0 |
| 25 | 3.229 | | 1.308 | 1.508 | | | | | -0.2 |
| 26 | 3.229 | | 1.096 | 1.508 | | | | | -0.412 |
| 27 | 3.002 | 3.002 | 2.154 | 1.531 | 1.531 | 1.471 | 0.06 | -0.03 | 0.623 |
| 28 | 3.002 | | 1.943 | 1.531 | | | | | 0.412 |
| 29 | 3.002 | | 1.731 | 1.531 | | | | | 0.2 |
| 30 | 3.264 | 3.264 | 2.097 | 1.638 | 1.638 | 1.626 | 0.012 | -0.01 | 0.459 |
| 31 | 2.965 | 2.965 | 1.141 | 1.365 | 1.365 | 1.6 | -0.235 | 0.13 | -0.224 |
| 32 | 2.965 | | 0.918 | 1.365 | | | | | -0.447 |
| 33 | 3.451 | 3.451 | 1.119 | 1.555 | 1.555 | 1.896 | -0.341 | 0.14 | -0.436 |
| 34 | 3.461 | 3.461 | 1.366 | 1.589 | 1.589 | 1.872 | -0.283 | 0.11 | -0.223 |
| 35 | 3.226 | 3.226 | 1.589 | 1.366 | 1.366 | 1.86 | -0.494 | 0.23 | 0.223 |
| 36 | 3.226 | | 1.366 | 1.366 | | | | | 0 |
| 37 | 3.2 | 3.2 | 1.341 | 1.341 | 1.341 | 1.859 | -0.518 | 0.25 | 0 |
| 38 | 3.2 | | 1.129 | 1.341 | | | | | -0.212 |
| 39 | 3.2 | | 0.918 | 1.341 | | | | | -0.423 |
| 40 | 2.981 | 2.981 | 1.59 | 1.367 | 1.367 | 1.614 | -0.247 | 0.13 | 0.223 |
| 41 | 3.494 | 3.494 | 2.318 | 1.659 | 1.659 | 1.835 | -0.176 | 0.07 | 0.659 |
| 42 | 3.494 | | 2.094 | 1.659 | | | | | 0.435 |
| 43 | 3.494 | | 1.882 | 1.659 | | | | | 0.223 |
| 44 | 3.494 | | 1.212 | 1.659 | | | | | -0.447 |
| 45 | 3.759 | 3.759 | 2.758 | 2.003 | 2.003 | 1.756 | 0.247 | -0.08 | 0.755 |
| 46 | 3.759 | | 2.487 | 2.003 | | | | | 0.484 |
| 47 | 3.759 | | 2.239 | 2.003 | | | | | 0.236 |
| 48 | 3.777 | 3.777 | 2 | 2 | 2 | 1.777 | 0.223 | -0.08 | 0 |
| 49 | 2.748 | 2.748 | 2.475 | 1.745 | 1.745 | 1.003 | 0.742 | -0.51 | 0.73 |
| 50 | 2.748 | | 1.98 | 1.745 | | | | | 0.235 |
| 51 | 3.697 | 3.697 | 1.26 | 1.719 | 1.719 | 1.978 | -0.259 | 0.09 | -0.459 |
| 52 | 3.475 | 3.475 | 1.437 | 1.437 | 1.437 | 2.038 | -0.601 | 0.25 | 0 |
| 53 | 3.389 | 3.389 | 1.6 | 1.6 | 1.6 | 1.789 | -0.189 | 0.08 | 0 |
| 54 | 2.942 | 2.942 | 1.789 | 1.353 | 1.353 | 1.589 | -0.236 | 0.13 | 0.436 |
| 55 | 2.942 | | 1.153 | 1.353 | | | | | -0.2 |
| 56 | 2.942 | | 0.93 | 1.353 | | | | | -0.423 |
| 57 | 3.204 | 3.204 | 2.25 | 1.567 | 1.567 | 1.637 | -0.07 | 0.03 | 0.683 |
| 58 | 3.192 | 3.192 | 1.19 | 1.366 | 1.366 | 1.826 | -0.46 | 0.22 | -0.176 |
| 59 | 3.214 | 3.214 | 1.754 | 1.754 | 1.754 | 1.46 | 0.294 | -0.14 | 0 |
| 60 | 3.194 | 3.194 | 2.145 | 1.521 | 1.521 | 1.673 | -0.152 | 0.07 | 0.624 |
| 61 | 3.194 | | 1.945 | 1.521 | | | | | 0.424 |
| 62 | 3.194 | | 1.733 | 1.521 | | | | | 0.212 |
| 63 | 3.18 | 3.18 | 1.531 | 1.531 | 1.531 | 1.649 | -0.118 | 0.06 | 0 |
| 64 | 3.18 | | 1.319 | 1.531 | | | | | -0.212 |
| 65 | 3.19 | 3.19 | 2.378 | 1.766 | 1.766 | 1.424 | 0.342 | -0.16 | 0.612 |
| 66 | 3.19 | | 2.178 | 1.766 | | | | | 0.412 |
| 67 | 3.19 | | 1.966 | 1.766 | | | | | 0.2 |
| 68 | 3.252 | 3.252 | 2.18 | 1.756 | 1.756 | 1.496 | 0.26 | -0.12 | 0.424 |
| 69 | 3.252 | | 1.756 | 1.756 | | | | | 0 |
| 70 | 3.287 | 3.287 | 2.639 | 1.991 | 1.991 | 1.296 | 0.695 | -0.32 | 0.648 |
| 71 | 3.287 | | 2.427 | 1.991 | | | | | 0.436 |
| 72 | 3.287 | | 1.991 | 1.991 | | | | | 0 |
| 73 | 3.287 | | 1.779 | 1.991 | | | | | -0.212 |
| 74 | 3.204 | 3.204 | 1.343 | 1.556 | 1.556 | 1.648 | -0.092 | 0.04 | -0.213 |
| 75 | 3.191 | 3.191 | 1.778 | 1.778 | 1.778 | 1.413 | 0.365 | -0.18 | 0 |
| 76 | 3.018 | 3.018 | 1.556 | 1.556 | 1.556 | 1.462 | 0.094 | -0.05 | 0 |
| 77 | 3.018 | | 1.332 | 1.556 | | | | | -0.224 |
| 78 | 3.018 | | 1.108 | 1.556 | | | | | -0.448 |
| 79 | 3.06 | 3.06 | 2.178 | 1.742 | 1.742 | 1.318 | 0.424 | -0.22 | 0.436 |
| 80 | 3.06 | | 1.954 | 1.742 | | | | | 0.212 |
| 81 | 3.178 | 3.178 | 1.954 | 1.754 | 1.754 | 1.424 | 0.33 | -0.16 | 0.2 |
| 82 | 3.027 | 3.027 | 2.391 | 1.731 | 1.731 | 1.296 | 0.435 | -0.23 | 0.66 |
| 83 | 3.049 | 3.049 | 1.966 | 1.542 | 1.542 | 1.507 | 0.035 | -0.02 | 0.424 |
| 84 | 3.049 | | 1.319 | 1.542 | | | | | -0.223 |
| 85 | 3.061 | 3.061 | 2.19 | 1.531 | 1.531 | 1.53 | 0.001 | 0.00 | 0.659 |
| 86 | 3.061 | | 1.978 | 1.531 | | | | | 0.447 |
| 87 | 3.061 | | 1.766 | 1.531 | | | | | 0.235 |
| 88 | 3.487 | 3.487 | 1.543 | 1.543 | 1.543 | 1.944 | -0.401 | 0.16 | 0 |
| 89 | 3.487 | | 1.319 | 1.543 | | | | | -0.224 |
| 90 | 3.04 | 3.04 | 1.98 | 1.756 | 1.756 | 1.284 | 0.472 | -0.25 | 0.224 |
| 91 | 3.04 | | 1.756 | 1.756 | | | | | 0 |
| 92 | 3.236 | 3.236 | 2.118 | 1.494 | 1.494 | 1.742 | -0.248 | 0.11 | 0.624 |
| 93 | 3.109 | 3.109 | 2.45 | 1.567 | 1.567 | 1.542 | 0.025 | -0.01 | 0.883 |
| 94 | 3.109 | | 2.238 | 1.567 | | | | | 0.671 |
| 95 | 3.109 | | 2.014 | 1.567 | | | | | 0.447 |
| 96 | 3.123 | 3.123 | 1.355 | 1.579 | 1.579 | 1.544 | 0.035 | -0.02 | -0.224 |
| 97 | 3.123 | | 1.143 | 1.579 | | | | | -0.436 |
| 98 | 3.112 | 3.112 | 1.981 | 1.309 | 1.309 | 1.803 | -0.494 | 0.25 | 0.672 |
| 99 | 3.112 | | 1.768 | 1.309 | | | | | 0.459 |
| 100 | 3.112 | | 1.309 | 1.309 | | | | | 0 |
| 101 | 3.086 | 3.086 | 2.426 | 1.555 | 1.555 | 1.531 | 0.024 | -0.01 | 0.871 |
| 102 | 3.086 | | 2.203 | 1.555 | | | | | 0.648 |
| 103 | 3.336 | 3.336 | 1.98 | 1.568 | 1.568 | 1.768 | -0.2 | 0.09 | 0.412 |
| 104 | 3.336 | | 1.768 | 1.568 | | | | | 0.2 |
| 105 | 3.336 | | 1.568 | 1.568 | | | | | 0 |
| 106 | 3.336 | | 1.356 | 1.568 | | | | | -0.212 |
| 107 | 3.336 | | 1.155 | 1.568 | | | | | -0.413 |
| 108 | 3.383 | 3.383 | 1.556 | 1.556 | 1.556 | 1.827 | -0.271 | 0.11 | 0 |
| 109 | 3.348 | 3.348 | 1.992 | 1.356 | 1.356 | 1.992 | -0.636 | 0.28 | 0.636 |
| 110 | 3.348 | | 1.356 | 1.356 | | | | | 0 |
| 111 | 3.348 | | 1.143 | 1.356 | | | | | -0.213 |
| 112 | 3.348 | | 0.931 | 1.356 | | | | | -0.425 |
| 113 | 3.535 | 3.535 | 2.628 | 1.756 | 1.756 | 1.779 | -0.023 | 0.01 | 0.872 |
| 114 | 3.535 | | 2.404 | 1.756 | | | | | 0.648 |
| 115 | 3.535 | | 2.192 | 1.756 | | | | | 0.436 |
| 116 | 3.535 | | 1.956 | 1.756 | | | | | 0.2 |
| 117 | 3.107 | 3.107 | 1.33 | 1.33 | 1.33 | 1.777 | -0.447 | 0.23 | 0 |
| 118 | 3.107 | | 1.118 | 1.33 | | | | | -0.212 |
| 119 | 3.107 | | 0.894 | 1.33 | | | | | -0.436 |
| 120 | 2.862 | 2.862 | 1.743 | 1.107 | 1.107 | 1.755 | -0.648 | 0.40 | 0.636 |
| 121 | 2.862 | | 1.519 | 1.107 | | | | | 0.412 |
| 122 | 2.862 | | 1.107 | 1.107 | | | | | 0 |
| 123 | 3.334 | 3.334 | 2.427 | 2.015 | 2.015 | 1.319 | 0.696 | -0.32 | 0.412 |
| 124 | 3.334 | | 2.215 | 2.015 | | | | | 0.2 |
| 125 | 3.334 | | 2.015 | 2.015 | | | | | 0 |
| 126 | 3.098 | 3.098 | 1.119 | 1.331 | 1.331 | 1.767 | -0.436 | 0.22 | -0.212 |
| 127 | 3.098 | | 0.895 | 1.331 | | | | | -0.436 |
| 128 | 3.307 | 3.307 | 1.801 | 1.801 | 1.801 | 1.506 | 0.295 | -0.13 | 0 |
| 129 | 3.307 | | 1.577 | 1.801 | | | | | -0.224 |
| 130 | 3.275 | 3.275 | 2.18 | 1.532 | 1.532 | 1.743 | -0.211 | 0.10 | 0.648 |
| 131 | 3.275 | | 1.744 | 1.532 | | | | | 0.212 |
| 132 | 3.275 | | 1.32 | 1.532 | | | | | -0.212 |
| 133 | 3.275 | | 1.107 | 1.532 | | | | | -0.425 |
| 134 | 3.298 | 3.298 | 1.084 | 1.496 | 1.496 | 1.802 | -0.306 | 0.14 | -0.412 |
| 135 | 3.039 | 3.039 | 1.932 | 1.519 | 1.519 | 1.52 | -0.001 | 0.00 | 0.413 |
| 136 | 3.039 | | 1.731 | 1.519 | | | | | 0.212 |
| 137 | 3.039 | | 1.519 | 1.519 | | | | | 0 |
| 138 | 3.039 | | 1.107 | 1.519 | | | | | -0.412 |
| 139 | 3.247 | 3.247 | 1.741 | 1.529 | 1.529 | 1.718 | -0.189 | 0.09 | 0.212 |
| 140 | 3.247 | | 1.529 | 1.529 | | | | | 0 |
| 141 | 3.247 | | 1.318 | 1.529 | | | | | -0.211 |
| 142 | 3.253 | 3.253 | 2.157 | 1.544 | 1.544 | 1.709 | -0.165 | 0.08 | 0.613 |
| 143 | 3.253 | | 1.956 | 1.544 | | | | | 0.412 |
| 144 | 3.253 | | 1.544 | 1.544 | | | | | 0 |
| 145 | 3.253 | | 1.343 | 1.544 | | | | | -0.201 |
| 146 | 3.253 | | 1.131 | 1.544 | | | | | -0.413 |
| 147 | 3.309 | 3.309 | 1.531 | 1.531 | 1.531 | 1.778 | -0.247 | 0.11 | 0 |
| 148 | 3.309 | | 1.307 | 1.531 | | | | | -0.224 |
| 149 | 3.25 | 3.25 | 1.507 | 1.719 | 1.719 | 1.531 | 0.188 | -0.09 | -0.212 |
| 150 | 3.192 | 3.192 | 1.59 | 1.59 | 1.59 | 1.602 | -0.012 | 0.01 | 0 |
| 151 | 3.641 | 3.641 | 1.591 | 1.591 | 1.591 | 2.05 | -0.459 | 0.17 | 0 |
| 152 | 3.641 | | 1.367 | 1.591 | | | | | -0.224 |
| 153 | 3.654 | 3.654 | 2.275 | 1.615 | 1.615 | 2.039 | -0.424 | 0.16 | 0.66 |
| 154 | 3.415 | 3.415 | 2.061 | 1.86 | 1.86 | 1.555 | 0.305 | -0.13 | 0.201 |
| 155 | 3.415 | | 1.649 | 1.86 | | | | | -0.211 |
| 156 | 3.167 | 3.167 | 1.542 | 1.542 | 1.542 | 1.625 | -0.083 | 0.04 | 0 |
| 157 | 3.179 | 3.179 | 0.907 | 1.542 | 1.542 | 1.637 | -0.095 | 0.05 | -0.635 |
| 158 | 3.194 | 3.194 | 1.815 | 1.591 | 1.591 | 1.603 | -0.012 | 0.01 | 0.224 |
| 159 | 3.194 | | 1.591 | 1.591 | | | | | 0 |
| 160 | 3.194 | | 1.379 | 1.591 | | | | | -0.212 |
| 161 | 3.428 | 3.428 | 2.273 | 1.826 | 1.826 | 1.602 | 0.224 | -0.09 | 0.447 |
| 162 | 3.189 | 3.189 | 1.389 | 1.601 | 1.601 | 1.588 | 0.013 | -0.01 | -0.212 |
| 163 | 3.189 | | 1.177 | 1.601 | | | | | -0.424 |
| 164 | 2.943 | 2.943 | 1.589 | 1.377 | 1.377 | 1.566 | -0.189 | 0.11 | 0.212 |
| 165 | 2.943 | | 1.377 | 1.377 | | | | | 0 |
| 166 | 3.427 | 3.427 | 1.826 | 1.826 | 1.826 | 1.601 | 0.225 | -0.09 | 0 |
| 167 | 3.17 | 3.17 | 2.039 | 1.614 | 1.614 | 1.556 | 0.058 | -0.03 | 0.425 |
| 168 | 3.17 | | 1.815 | 1.614 | | | | | 0.201 |
| 169 | 2.966 | 2.966 | 1.601 | 1.354 | 1.354 | 1.612 | -0.258 | 0.14 | 0.247 |
| 170 | 2.966 | | 1.354 | 1.354 | | | | | 0 |
| 171 | 3.235 | 3.235 | 2.047 | 1.835 | 1.835 | 1.4 | 0.435 | -0.20 | 0.212 |
| 172 | 3.218 | 3.218 | 1.803 | 1.579 | 1.579 | 1.639 | -0.06 | 0.03 | 0.224 |
| 173 | 3.448 | 3.448 | 1.801 | 1.553 | 1.553 | 1.895 | -0.342 | 0.14 | 0.248 |
| 174 | 3.448 | | 1.553 | 1.553 | | | | | 0 |
| 175 | 3.235 | 3.235 | 2.035 | 1.565 | 1.565 | 1.67 | -0.105 | 0.05 | 0.47 |
| 176 | 3.235 | | 1.8 | 1.565 | | | | | 0.235 |
| 177 | 3.203 | 3.203 | 1.578 | 1.331 | 1.331 | 1.872 | -0.541 | 0.26 | 0.247 |
| 178 | 3.203 | | 1.331 | 1.331 | | | | | 0 |
| 179 | 3.167 | 3.167 | 1.154 | 1.601 | 1.601 | 1.566 | 0.035 | -0.02 | -0.447 |
| 180 | 3.424 | 3.424 | 2.742 | 1.859 | 1.859 | 1.565 | 0.294 | -0.12 | 0.883 |
| 181 | 3.424 | | 2.518 | 1.859 | | | | | 0.659 |
| 182 | 3.424 | | 2.071 | 1.859 | | | | | 0.212 |
| 183 | 3.424 | | 1.859 | 1.859 | | | | | 0 |
| 184 | 3.168 | 3.168 | 2.238 | 1.343 | 1.343 | 1.825 | -0.482 | 0.24 | 0.895 |
| 185 | 3.168 | | 2.014 | 1.343 | | | | | 0.671 |
| 186 | 3.168 | | 1.778 | 1.343 | | | | | 0.435 |
| 187 | 3.423 | 3.423 | 1.588 | 1.388 | 1.388 | 2.035 | -0.647 | 0.28 | 0.2 |
| 188 | 3.2 | 3.2 | 1.306 | 1.306 | 1.306 | 1.894 | -0.588 | 0.29 | 0 |
| 189 | 3.192 | 3.192 | 2.274 | 1.402 | 1.402 | 1.79 | -0.388 | 0.19 | 0.872 |
| 190 | 3.192 | | 2.05 | 1.402 | | | | | 0.648 |
| 191 | 3.192 | | 1.826 | 1.402 | | | | | 0.424 |
| 192 | 3.413 | 3.413 | 1.801 | 1.601 | 1.601 | 1.812 | -0.211 | 0.09 | 0.2 |
| 193 | 3.247 | 3.247 | 1.376 | 1.141 | 1.141 | 2.106 | -0.965 | 0.48 | 0.235 |
| 194 | 3.247 | | 1.141 | 1.141 | | | | | 0 |
| 195 | 3.454 | 3.454 | 1.874 | 1.874 | 1.874 | 1.58 | 0.294 | -0.12 | 0 |
| 196 | 3.451 | 3.451 | 1.178 | 1.414 | 1.414 | 2.037 | -0.623 | 0.26 | -0.236 |
| 197 | 3.419 | 3.419 | 1.379 | 1.603 | 1.603 | 1.816 | -0.213 | 0.09 | -0.224 |
| 198 | 3.235 | 3.235 | 1.365 | 1.153 | 1.153 | 2.082 | -0.929 | 0.47 | 0.212 |
| 199 | 3.235 | | 1.153 | 1.153 | | | | | 0 |
| 200 | 3.235 | | 0.929 | 1.153 | | | | | -0.224 |
| 201 | 3.247 | 3.247 | 1.847 | 1.141 | 1.141 | 2.106 | -0.965 | 0.48 | 0.706 |
| 202 | 3.247 | | 1.612 | 1.141 | | | | | 0.471 |
| 203 | 3.247 | | 1.377 | 1.141 | | | | | 0.236 |
| 204 | 3.247 | | 1.141 | 1.141 | | | | | 0 |
| 205 | 3.472 | 3.472 | 2.554 | 1.636 | 1.636 | 1.836 | -0.2 | 0.08 | 0.918 |
| 206 | 3.472 | | 2.319 | 1.636 | | | | | 0.683 |
| 207 | 3.472 | | 2.095 | 1.636 | | | | | 0.459 |
| 208 | 3.447 | 3.447 | 1.8 | 1.365 | 1.365 | 2.082 | -0.717 | 0.30 | 0.435 |
| 209 | 3.447 | | 1.365 | 1.365 | | | | | 0 |
| 210 | 3.229 | 3.229 | 1.178 | 1.614 | 1.614 | 1.615 | -0.001 | 0.00 | -0.436 |
| 211 | 2.969 | 2.969 | 1.402 | 1.637 | 1.637 | 1.332 | 0.305 | -0.17 | -0.235 |
| 212 | 2.969 | | 1.178 | 1.637 | | | | | -0.459 |
| 213 | 3.393 | 3.393 | 2.486 | 1.567 | 1.567 | 1.826 | -0.259 | 0.11 | 0.919 |
| 214 | 3.393 | | 2.25 | 1.567 | | | | | 0.683 |
| 215 | 3.404 | 3.404 | 2.05 | 1.826 | 1.826 | 1.578 | 0.248 | -0.10 | 0.224 |
| 216 | 3.404 | | 1.826 | 1.826 | | | | | 0 |
| 217 | 3.723 | 3.723 | 1.85 | 1.64 | 1.64 | 2.083 | -0.443 | 0.16 | 0.21 |
| 218 | 3.723 | | 1.64 | 1.64 | | | | | 0 |
| 219 | 3.723 | | 1.402 | 1.64 | | | | | -0.238 |
| 220 | 3.472 | 3.472 | 2.071 | 1.601 | 1.601 | 1.871 | -0.27 | 0.11 | 0.47 |
| 221 | 3.259 | 3.259 | 1.835 | 1.376 | 1.376 | 1.883 | -0.507 | 0.24 | 0.459 |
| 222 | 3.259 | | 1.612 | 1.376 | | | | | 0.236 |
| 223 | 3.271 | 3.271 | 1.376 | 1.165 | 1.165 | 2.106 | -0.941 | 0.46 | 0.211 |
| 224 | 3.271 | | 0.941 | 1.165 | | | | | -0.224 |
| 225 | 3.248 | 3.248 | 2.542 | 1.612 | 1.612 | 1.636 | -0.024 | 0.01 | 0.93 |
| 226 | 3.248 | | 2.318 | 1.612 | | | | | 0.706 |
| 227 | 3.248 | | 2.083 | 1.612 | | | | | 0.471 |
| 228 | 3.248 | | 1.847 | 1.612 | | | | | 0.235 |
| 229 | 3.461 | 3.461 | 1.848 | 2.072 | 2.072 | 1.389 | 0.683 | -0.29 | -0.224 |
| 230 | 3.201 | 3.201 | 2.048 | 1.812 | 1.812 | 1.389 | 0.423 | -0.20 | 0.236 |
| 231 | 2.993 | 2.993 | 1.39 | 1.39 | 1.39 | 1.603 | -0.213 | 0.12 | 0 |
| 232 | 3.016 | 3.016 | 1.178 | 1.637 | 1.637 | 1.379 | 0.258 | -0.14 | -0.459 |
| 233 | 3.52 | 3.52 | 1.613 | 1.613 | 1.613 | 1.907 | -0.294 | 0.12 | 0 |
| 234 | 3.448 | 3.448 | 1.86 | 1.86 | 1.86 | 1.588 | 0.272 | -0.11 | 0 |
| 235 | 3.448 | | 1.624 | 1.86 | | | | | -0.236 |
| 236 | 3.448 | | 1.401 | 1.86 | | | | | -0.459 |
| 237 | 3.217 | 3.217 | 1.815 | 1.379 | 1.379 | 1.838 | -0.459 | 0.22 | 0.436 |
| 238 | 3.217 | | 1.591 | 1.379 | | | | | 0.212 |
| 239 | 3.217 | | 1.155 | 1.379 | | | | | -0.224 |
| 240 | 3.461 | 3.461 | 1.813 | 1.577 | 1.577 | 1.884 | -0.307 | 0.12 | 0.236 |
| 241 | 3.461 | | 1.577 | 1.577 | | | | | 0 |
| | | | | | | | | | |
| | | | | | $d_{l(\mathrm{ave})}$ [nm] | $d_{u(\mathrm{ave})}$ [nm] | $d_{l\text{-}u(\mathrm{ave})}$ [nm] | $D_{\mathrm{net(ave)}}$ [mJ/m$^2$] | |
| | | | | | **1.569** | **1.688** | **-0.119** | **0.05** | |

Table S5. Numerical data of various distance measurements and number of layers for the 1.0-sample. The reason why half-integers are shown in $N_{Co}$ is that the middle of the adjacent Co-richest layers is defined as the virtual Co-richest layer. The sample numbers of Fig. 3(b) in the main text are rearranged in order of increasing $d_{l-u}$ value to make the dispersion easier to recognize.

| Sample number | Distance between richest Pt at top and bottom | Distance of the richest Co from the bottom richest Pt **[Fig. 3(b)]** | Distance of the richest Co from the top richest Pt **[Fig. 3(b)]** | Distance between $d_l$ and $d_u$ | Net Bulk DMI energy **[Fig. 3(d)]** | Number of layers from the bottom to top Pt-richest layer | Layer position of the Co-richest layer counting from the bottom Pt-richest layer |
|---|---|---|---|---|---|---|---|
| | $d_{Pt\text{-}Pt}$ [nm] | $d_l$ [nm] | $d_u$ [nm] | $d_{l-u}$ [nm] | $D_{net}$ [mJ/m$^2$] | $N_{Pt\text{-}Pt}$ [number] | $N_{Co}$ [number] |
| 1 | 2.599 | 1.092 | 1.506 | -0.414 | 0.30 | 13 | 6 |
| 2 | 2.806 | 1.280 | 1.525 | -0.245 | 0.15 | 14 | 7 |
| 3 | 2.617 | 1.064 | 1.554 | -0.490 | 0.36 | 13 | 6 |
| 4 | 2.843 | 1.337 | 1.506 | -0.169 | 0.10 | 14 | 7 |
| 5 | 2.843 | 1.337 | 1.506 | -0.169 | 0.10 | 14 | 7 |
| 6 | 2.843 | 1.318 | 1.525 | -0.207 | 0.12 | 14 | 7 |
| 7 | 2.655 | 0.904 | 1.544 | -0.640 | 0.55 | 13 | 5.5 |
| 8 | 2.617 | 1.101 | 1.515 | -0.414 | 0.30 | 13 | 6 |
| 9 | 2.645 | 1.148 | 1.497 | -0.348 | 0.24 | 13 | 6 |
| 10 | 3.012 | 1.271 | 1.741 | -0.470 | 0.26 | 15 | 7 |
| 11 | 3.012 | 1.252 | 1.760 | -0.508 | 0.28 | 15 | 7 |
| 12 | 2.579 | 1.064 | 1.515 | -0.452 | 0.34 | 13 | 6 |
| 13 | 2.976 | 1.394 | 1.582 | -0.188 | 0.10 | 14 | 7 |
| 14 | 2.665 | 1.318 | 1.347 | -0.028 | 0.02 | 13 | 7 |
| 15 | 3.116 | 1.337 | 1.779 | -0.443 | 0.22 | 15 | 7 |
| 16 | 3.032 | 1.497 | 1.535 | -0.038 | 0.02 | 15 | 8 |
| 17 | 3.182 | 1.798 | 1.384 | 0.414 | -0.20 | 15 | 9 |
| 18 | 3.107 | 1.544 | 1.563 | -0.019 | 0.01 | 15 | 8 |
| 19 | 2.871 | 1.732 | 1.139 | 0.593 | -0.36 | 14 | 9 |
| 20 | 3.154 | 1.318 | 1.601 | -0.283 | 0.16 | 15 | 7.5 |
| 21 | 3.136 | 1.507 | 1.629 | -0.123 | 0.06 | 15 | 8 |
| 22 | 3.117 | 1.507 | 1.610 | -0.104 | 0.05 | 15 | 8 |
| 23 | 2.872 | 1.525 | 1.346 | 0.179 | -0.11 | 14 | 8 |
| 24 | 2.871 | 1.525 | 1.346 | 0.179 | -0.11 | 14 | 8 |
| 25 | 3.069 | 1.516 | 1.553 | -0.038 | 0.02 | 15 | 8 |
| 26 | 2.890 | 1.572 | 1.318 | 0.254 | -0.15 | 14 | 8 |
| 27 | 3.333 | 1.742 | 1.591 | 0.151 | -0.07 | 16 | 9 |
| 28 | 2.890 | 1.299 | 1.591 | -0.292 | 0.17 | 14 | 7 |
| 29 | 2.928 | 1.563 | 1.365 | 0.198 | -0.11 | 14 | 8 |
| 30 | 2.890 | 1.534 | 1.356 | 0.179 | -0.10 | 14 | 8 |
| 31 | 2.881 | 1.553 | 1.327 | 0.226 | -0.13 | 14 | 8 |
| 32 | 3.078 | 1.534 | 1.544 | -0.010 | 0.00 | 15 | 8 |
| 33 | 2.908 | 1.337 | 1.327 | 0.010 | -0.01 | 14 | 7.5 |
| 34 | 2.918 | 1.177 | 1.544 | -0.367 | 0.24 | 14 | 6.5 |
| 35 | 3.022 | 1.440 | 1.365 | 0.075 | -0.05 | 14 | 7.5 |
| 36 | 3.136 | 1.591 | 1.544 | 0.047 | -0.02 | 15 | 8 |
| 37 | 2.776 | 1.619 | 1.158 | 0.461 | -0.30 | 13 | 8 |
| 38 | 2.899 | 1.798 | 1.101 | 0.696 | -0.42 | 14 | 9 |
| 39 | 3.096 | 1.562 | 1.299 | 0.264 | -0.16 | 15 | 8.5 |
| 40 | 2.937 | 1.581 | 1.355 | 0.226 | -0.13 | 14 | 8 |
| 41 | 2.918 | 1.346 | 1.572 | -0.226 | 0.13 | 14 | 7 |
| 42 | 2.664 | 1.111 | 1.337 | -0.226 | 0.18 | 13 | 6.5 |
| 43 | 2.654 | 1.120 | 1.534 | -0.414 | 0.29 | 13 | 6 |
| 44 | 2.918 | 1.129 | 1.788 | -0.659 | 0.39 | 14 | 6 |
| 45 | 2.927 | 1.402 | 1.525 | -0.122 | 0.07 | 14 | 7 |
| 46 | 2.937 | 1.374 | 1.562 | -0.188 | 0.11 | 14 | 7 |
| 47 | 2.918 | 1.356 | 1.563 | -0.207 | 0.12 | 14 | 7 |
| 48 | 2.721 | 1.374 | 1.346 | 0.028 | -0.02 | 13 | 7 |
| 49 | 3.154 | 1.394 | 1.761 | -0.367 | 0.18 | 15 | 7 |
| 50 | 3.145 | 1.563 | 1.356 | 0.207 | -0.12 | 15 | 8.5 |
| 51 | 2.721 | 1.375 | 1.111 | 0.264 | -0.21 | 13 | 7.5 |
| 52 | 2.712 | 1.346 | 1.130 | 0.217 | -0.17 | 13 | 7.5 |
| 53 | 2.712 | 1.347 | 1.140 | 0.207 | -0.16 | 13 | 7.5 |
| 54 | 2.467 | 0.876 | 1.592 | -0.716 | 0.62 | 12 | 5 |
| 55 | 2.899 | 1.082 | 1.544 | -0.461 | 0.33 | 14 | 6.5 |
| 56 | 2.443 | 1.120 | 1.087 | 0.033 | -0.03 | 12 | 6.5 |
| 57 | 2.890 | 1.562 | 1.327 | 0.235 | -0.14 | 14 | 8 |
| 58 | 2.927 | 1.591 | 1.337 | 0.254 | -0.14 | 14 | 8 |
| 59 | 2.976 | 1.799 | 1.177 | 0.622 | -0.35 | 14 | 9 |
| 60 | 2.948 | 1.582 | 1.121 | 0.462 | -0.31 | 14 | 8.5 |
| 61 | 3.004 | 1.620 | 1.121 | 0.499 | -0.33 | 14 | 8.5 |
| 62 | 2.656 | 1.601 | 1.055 | 0.546 | -0.39 | 13 | 8 |
| 63 | 3.136 | 1.591 | 1.544 | 0.047 | -0.02 | 15 | 8 |
| 64 | 3.146 | 1.582 | 1.347 | 0.235 | -0.13 | 15 | 8.5 |
| 65 | 2.929 | 1.356 | 1.356 | 0.000 | 0.00 | 14 | 7.5 |
| 66 | 2.843 | 1.346 | 1.280 | 0.066 | -0.05 | 14 | 7.5 |
| 67 | 2.862 | 1.158 | 1.506 | -0.348 | 0.24 | 14 | 6.5 |
| 68 | 2.401 | 1.328 | 0.885 | 0.443 | -0.45 | 12 | 7.5 |
| 69 | 2.844 | 1.102 | 1.742 | -0.640 | 0.40 | 14 | 6 |
| 70 | 3.098 | 1.592 | 1.507 | 0.085 | -0.04 | 15 | 8 |
| 71 | 2.891 | 1.356 | 1.535 | -0.179 | 0.10 | 14 | 7 |
| 72 | 3.040 | 1.751 | 1.289 | 0.461 | -0.25 | 15 | 9 |
| 73 | 2.861 | 1.299 | 1.562 | -0.264 | 0.16 | 14 | 7 |
| 74 | 3.332 | 1.572 | 1.760 | -0.188 | 0.08 | 16 | 8 |
| 75 | 2.861 | 1.327 | 1.308 | 0.019 | -0.01 | 14 | 7.5 |
| 76 | 2.618 | 1.347 | 1.271 | 0.075 | -0.05 | 13 | 7 |
| 77 | 3.249 | 1.733 | 1.516 | 0.217 | -0.10 | 16 | 9 |
| 78 | 2.655 | 1.328 | 1.111 | 0.217 | -0.18 | 13 | 7.5 |
| 79 | 3.324 | 1.554 | 1.563 | -0.009 | 0.00 | 16 | 8.5 |
| 80 | 2.927 | 1.365 | 1.563 | -0.198 | 0.11 | 14 | 7 |
| 81 | 3.116 | 1.553 | 1.337 | 0.216 | -0.13 | 15 | 8.5 |
| 82 | 2.673 | 1.111 | 1.346 | -0.235 | 0.19 | 13 | 6.5 |
| 83 | 3.379 | 1.581 | 1.591 | -0.009 | 0.00 | 16 | 8.5 |
| 84 | 3.408 | 1.883 | 1.346 | 0.537 | -0.26 | 16 | 9.5 |
| 85 | 2.485 | 1.158 | 1.111 | 0.047 | -0.04 | 12 | 6.5 |
| 86 | 2.946 | 1.600 | 1.346 | 0.254 | -0.14 | 14 | 8 |
| 87 | 3.135 | 1.591 | 1.327 | 0.264 | -0.15 | 15 | 8.5 |
| 88 | 2.937 | 1.365 | 1.365 | 0.000 | 0.00 | 14 | 7.5 |
| 89 | 3.013 | 1.375 | 1.638 | -0.264 | 0.14 | 14 | 7 |
| 90 | 3.098 | 1.460 | 1.394 | 0.066 | -0.04 | 14 | 7.5 |
| 91 | 3.013 | 1.441 | 1.573 | -0.132 | 0.07 | 14 | 7 |
| 92 | 2.805 | 1.205 | 1.600 | -0.395 | 0.25 | 13 | 6 |
| 93 | 3.248 | 1.648 | 1.600 | 0.047 | -0.02 | 15 | 8 |
| 94 | 3.003 | 1.638 | 1.365 | 0.273 | -0.15 | 14 | 8 |
| 95 | 2.975 | 1.422 | 1.327 | 0.094 | -0.06 | 14 | 7.5 |
| 96 | 2.984 | 1.167 | 1.572 | -0.405 | 0.27 | 14 | 6.5 |
| 97 | 3.013 | 1.873 | 1.139 | 0.734 | -0.41 | 14 | 9 |
| 98 | 2.523 | 0.932 | 1.346 | -0.414 | 0.40 | 12 | 5.5 |
| 99 | 2.542 | 1.158 | 1.384 | -0.226 | 0.17 | 12 | 6 |
| 100 | 3.012 | 1.657 | 1.355 | 0.301 | -0.16 | 14 | 8 |
| 101 | 2.562 | 1.187 | 1.375 | -0.188 | 0.14 | 12 | 6 |
| 102 | 2.298 | 0.951 | 1.111 | -0.160 | 0.18 | 11 | 5.5 |
| 103 | 3.249 | 1.158 | 1.846 | -0.688 | 0.39 | 15 | 6.5 |
| 104 | 2.995 | 1.394 | 1.601 | -0.207 | 0.11 | 14 | 7 |
| 105 | 2.505 | 1.337 | 1.168 | 0.170 | -0.13 | 12 | 7 |
| 106 | 2.806 | 1.177 | 1.384 | -0.207 | 0.15 | 13 | 6.5 |
| 107 | 2.730 | 1.563 | 1.167 | 0.396 | -0.26 | 13 | 8 |
| 108 | 2.777 | 1.149 | 1.384 | -0.235 | 0.18 | 13 | 6.5 |
| 109 | 2.909 | 1.365 | 1.544 | -0.179 | 0.10 | 14 | 7 |
| 110 | 2.946 | 1.393 | 1.318 | 0.075 | -0.05 | 14 | 7.5 |
| 111 | 2.975 | 1.167 | 1.807 | -0.640 | 0.37 | 14 | 6 |
| 112 | 2.946 | 1.158 | 1.789 | -0.631 | 0.37 | 14 | 6 |
| 113 | 2.956 | 1.836 | 1.120 | 0.715 | -0.42 | 14 | 9 |
| 114 | 3.107 | 1.516 | 1.346 | 0.169 | -0.10 | 15 | 8.5 |
| 115 | 2.900 | 1.290 | 1.365 | -0.075 | 0.05 | 14 | 7.5 |
| 116 | 3.135 | 1.309 | 1.582 | -0.273 | 0.16 | 15 | 7.5 |
| 117 | 2.899 | 1.337 | 1.563 | -0.226 | 0.13 | 14 | 7 |
| 118 | 2.946 | 1.139 | 1.591 | -0.452 | 0.30 | 14 | 6.5 |
| 119 | 2.730 | 1.120 | 1.384 | -0.264 | 0.20 | 13 | 6.5 |
| 120 | 2.683 | 1.111 | 1.337 | -0.226 | 0.18 | 13 | 6.5 |
| 121 | 2.494 | 1.130 | 1.130 | 0.000 | 0.00 | 12 | 6.5 |
| 122 | 2.946 | 1.403 | 1.544 | -0.141 | 0.08 | 14 | 7 |
| | | | | | | | |
| | | $d_{l(\mathrm{ave})}$ [nm] | $d_{u(\mathrm{ave})}$ [nm] | $d_{l\text{-}u(\mathrm{ave})}$ [nm] | $D_{\mathrm{net(ave)}}$ [mJ/m$^2$] | | |
| | | **1.388** | **1.427** | **-0.039** | **0.03** | | |

Table S6. Numerical data of various numbers of layers for the 0.4-sample, corresponding to Table S4. Deletion of duplicates is due to the presence of two or more dislocations in a single line profile. The sample numbers of Fig. 4(e) in the main text are rearranged in order of increasing number of $N_{d\text{-Co}}$ to make the distribution of the dislocation layers relative to the Co-richest layer easier to recognize.

| Sample number | Number of layers from the bottom to top Pt-richest layers | Remove duplicates from the left column | Layer position of the Co-richest layer counting from the bottom Pt-richest layer | Remove duplicates from the left column | Layer position of the dislocation layer counting from the bottom Pt-richest layer | Layer difference between $N_d$ and $N_{Co}$ **[Fig. 4(e)]** |
|---|---|---|---|---|---|---|
| | $N_{Pt\text{-}Pt}$ [number] | $N_{Pt\text{-}Pt}$ [number] | $N_{Co}$ [number] | $N_{Co}$ [number] | $N_d$ [number] | $N_{d\text{-Co}}$ [number] |
| 1 | 15 | 15 | 9 | 9 | 11 | 2 |
| 2 | 15 | | 9 | | 10 | 1 |
| 3 | 16 | 16 | 9 | 9 | 8 | -1 |
| 4 | 16 | | 9 | | 7 | -2 |
| 5 | 16 | 16 | 9 | 9 | 13 | 4 |
| 6 | 16 | | 9 | | 12 | 3 |
| 7 | 17 | 17 | 8 | 8 | 9 | 1 |
| 8 | 17 | 17 | 8 | 8 | 11 | 3 |
| 9 | 17 | | 8 | | 10 | 2 |
| 10 | 17 | | 8 | | 9 | 1 |
| 11 | 17 | | 8 | | 8 | 0 |
| 12 | 17 | | 8 | | 7 | -1 |
| 13 | 16 | 16 | 8 | 8 | 11 | 3 |
| 14 | 16 | | 8 | | 10 | 2 |
| 15 | 16 | | 8 | | 9 | 1 |
| 16 | 15 | 15 | 7 | 7 | 8 | 1 |
| 17 | 15 | | 7 | | 7 | 0 |
| 18 | 15 | | 7 | | 6 | -1 |
| 19 | 15 | 15 | 7 | 7 | 7 | 0 |
| 20 | 15 | | 7 | | 6 | -1 |
| 21 | 16 | 16 | 8 | 8 | 13 | 5 |
| 22 | 16 | | 8 | | 12 | 4 |
| 23 | 16 | | 8 | | 10 | 2 |
| 24 | 16 | | 8 | | 8 | 0 |
| 25 | 16 | | 8 | | 7 | -1 |
| 26 | 16 | | 8 | | 6 | -2 |
| 27 | 15 | 15 | 8 | 8 | 11 | 3 |
| 28 | 15 | | 8 | | 10 | 2 |
| 29 | 15 | | 8 | | 9 | 1 |
| 30 | 15 | 15 | 8 | 8 | 10 | 2 |
| 31 | 14 | 14 | 7 | 7 | 6 | -1 |
| 32 | 14 | | 7 | | 5 | -2 |
| 33 | 16 | 16 | 8 | 8 | 6 | -2 |
| 34 | 16 | 16 | 8 | 8 | 7 | -1 |
| 35 | 15 | 15 | 7 | 7 | 8 | 1 |
| 36 | 15 | | 7 | | 7 | 0 |
| 37 | 15 | 15 | 7 | 7 | 7 | 0 |
| 38 | 15 | | 7 | | 6 | -1 |
| 39 | 15 | | 7 | | 5 | -2 |
| 40 | 14 | 14 | 7 | 7 | 8 | 1 |
| 41 | 16 | 16 | 8 | 8 | 11 | 3 |
| 42 | 16 | | 8 | | 10 | 2 |
| 43 | 16 | | 8 | | 9 | 1 |
| 44 | 16 | | 8 | | 6 | -2 |
| 45 | 16 | 16 | 9 | 9 | 12 | 3 |
| 46 | 16 | | 9 | | 11 | 2 |
| 47 | 16 | | 9 | | 10 | 1 |
| 48 | 16 | 16 | 9 | 9 | 9 | 0 |
| 49 | 16 | 16 | 8 | 8 | 11 | 3 |
| 50 | 16 | | 8 | | 9 | 1 |
| 51 | 16 | 16 | 8 | 8 | 6 | -2 |
| 52 | 15 | 15 | 7 | 7 | 7 | 0 |
| 53 | 16 | 16 | 8 | 8 | 8 | 0 |
| 54 | 14 | 14 | 7 | 7 | 9 | 2 |
| 55 | 14 | | 7 | | 6 | -1 |
| 56 | 14 | | 7 | | 5 | -2 |
| 57 | 15 | 15 | 8 | 8 | 11 | 3 |
| 58 | 15 | 15 | 7 | 7 | 6 | -1 |
| 59 | 16 | 16 | 9 | 9 | 9 | 0 |
| 60 | 16 | 16 | 8 | 8 | 11 | 3 |
| 61 | 16 | | 8 | | 10 | 2 |
| 62 | 16 | | 8 | | 9 | 1 |
| 63 | 16 | 16 | 8 | 8 | 8 | 0 |
| 64 | 16 | | 8 | | 7 | -1 |
| 65 | 16 | 16 | 9 | 9 | 12 | 3 |
| 66 | 16 | | 9 | | 11 | 2 |
| 67 | 16 | | 9 | | 10 | 1 |
| 68 | 16 | 16 | 9 | 9 | 11 | 2 |
| 69 | 16 | | 9 | | 9 | 0 |
| 70 | 16 | 16 | 10 | 10 | 13 | 3 |
| 71 | 16 | | 10 | | 12 | 2 |
| 72 | 16 | | 10 | | 10 | 0 |
| 73 | 16 | | 10 | | 9 | -1 |
| 74 | 16 | 16 | 8 | 8 | 7 | -1 |
| 75 | 16 | 16 | 9 | 9 | 9 | 0 |
| 76 | 15 | 15 | 8 | 8 | 8 | 0 |
| 77 | 15 | | 8 | | 7 | -1 |
| 78 | 15 | | 8 | | 6 | -2 |
| 79 | 15 | 15 | 9 | 9 | 11 | 2 |
| 80 | 15 | | 9 | | 10 | 1 |
| 81 | 16 | 16 | 9 | 9 | 10 | 1 |
| 82 | 15 | 15 | 9 | 9 | 12 | 3 |
| 83 | 15 | 15 | 8 | 8 | 10 | 2 |
| 84 | 15 | | 8 | | 7 | -1 |
| 85 | 15 | 15 | 8 | 8 | 11 | 3 |
| 86 | 15 | | 8 | | 10 | 2 |
| 87 | 15 | | 8 | | 9 | 1 |
| 88 | 17 | 17 | 8 | 8 | 8 | 0 |
| 89 | 17 | | 8 | | 7 | -1 |
| 90 | 15 | 15 | 9 | 9 | 10 | 1 |
| 91 | 15 | | 9 | | 9 | 0 |
| 92 | 16 | 16 | 8 | 8 | 11 | 3 |
| 93 | 15 | 15 | 8 | 8 | 12 | 4 |
| 94 | 15 | | 8 | | 11 | 3 |
| 95 | 15 | | 8 | | 10 | 2 |
| 96 | 15 | 15 | 8 | 8 | 7 | -1 |
| 97 | 15 | | 8 | | 6 | -2 |
| 98 | 15 | 15 | 7 | 7 | 10 | 3 |
| 99 | 15 | | 7 | | 9 | 2 |
| 100 | 15 | | 7 | | 7 | 0 |
| 101 | 15 | 15 | 8 | 8 | 12 | 4 |
| 102 | 15 | | 8 | | 11 | 3 |
| 103 | 16 | 16 | 8 | 8 | 10 | 2 |
| 104 | 16 | | 8 | | 9 | 1 |
| 105 | 16 | | 8 | | 8 | 0 |
| 106 | 16 | | 8 | | 7 | -1 |
| 107 | 16 | | 8 | | 6 | -2 |
| 108 | 16 | 16 | 8 | 8 | 8 | 0 |
| 109 | 16 | 16 | 7 | 7 | 10 | 3 |
| 110 | 16 | | 7 | | 7 | 0 |
| 111 | 16 | | 7 | | 6 | -1 |
| 112 | 16 | | 7 | | 5 | -2 |
| 113 | 17 | 17 | 9 | 9 | 13 | 4 |
| 114 | 17 | | 9 | | 12 | 3 |
| 115 | 17 | | 9 | | 11 | 2 |
| 116 | 17 | | 9 | | 10 | 1 |
| 117 | 15 | 15 | 7 | 7 | 7 | 0 |
| 118 | 15 | | 7 | | 6 | -1 |
| 119 | 15 | | 7 | | 5 | -2 |
| 120 | 14 | 14 | 6 | 6 | 9 | 3 |
| 121 | 14 | | 6 | | 8 | 2 |
| 122 | 14 | | 6 | | 6 | 0 |
| 123 | 16 | 16 | 10 | 10 | 12 | 2 |
| 124 | 16 | | 10 | | 11 | 1 |
| 125 | 16 | | 10 | | 10 | 0 |
| 126 | 15 | 15 | 7 | 7 | 6 | -1 |
| 127 | 15 | | 7 | | 5 | -2 |
| 128 | 16 | 16 | 9 | 9 | 9 | 0 |
| 129 | 16 | | 9 | | 8 | -1 |
| 130 | 16 | 16 | 8 | 8 | 11 | 3 |
| 131 | 16 | | 8 | | 9 | 1 |
| 132 | 16 | | 8 | | 7 | -1 |
| 133 | 16 | | 8 | | 6 | -2 |
| 134 | 16 | 16 | 8 | 8 | 6 | -2 |
| 135 | 15 | 15 | 8 | 8 | 10 | 2 |
| 136 | 15 | | 8 | | 9 | 1 |
| 137 | 15 | | 8 | | 8 | 0 |
| 138 | 15 | | 8 | | 6 | -2 |
| 139 | 16 | 16 | 8 | 8 | 9 | 1 |
| 140 | 16 | | 8 | | 8 | 0 |
| 141 | 16 | | 8 | | 7 | -1 |
| 142 | 16 | 16 | 8 | 8 | 11 | 3 |
| 143 | 16 | | 8 | | 10 | 2 |
| 144 | 16 | | 8 | | 8 | 0 |
| 145 | 16 | | 8 | | 7 | -1 |
| 146 | 16 | | 8 | | 6 | -2 |
| 147 | 16 | 16 | 8 | 8 | 8 | 0 |
| 148 | 16 | | 8 | | 7 | -1 |
| 149 | 16 | 16 | 9 | 9 | 8 | -1 |
| 150 | 15 | 15 | 8 | 8 | 8 | 0 |
| 151 | 17 | 17 | 8 | 8 | 8 | 0 |
| 152 | 17 | | 8 | | 7 | -1 |
| 153 | 17 | 17 | 8 | 8 | 11 | 3 |
| 154 | 16 | 16 | 9 | 9 | 10 | 1 |
| 155 | 16 | | 9 | | 8 | |
| 156 | 15 | 15 | 8 | 8 | 8 | 0 |
| 157 | 15 | 15 | 8 | 8 | 5 | -3 |
| 158 | 15 | 15 | 8 | 8 | 9 | 1 |
| 159 | 15 | | 8 | | 8 | 0 |
| 160 | 15 | | 8 | | 7 | -1 |
| 161 | 16 | 16 | 9 | 9 | 11 | 2 |
| 162 | 15 | 15 | 8 | 8 | 7 | -1 |
| 163 | 15 | | 8 | | 6 | -2 |
| 164 | 14 | 14 | 7 | 7 | 8 | 1 |
| 165 | 14 | | 7 | | 7 | 0 |
| 166 | 16 | 16 | 9 | 9 | 9 | 0 |
| 167 | 15 | 15 | 8 | 8 | 10 | 2 |
| 168 | 15 | | 8 | | 9 | 1 |
| 169 | 14 | 14 | 7 | 7 | 8 | 1 |
| 170 | 14 | | 7 | | 7 | 0 |
| 171 | 15 | 15 | 9 | 9 | 10 | 1 |
| 172 | 15 | 15 | 8 | 8 | 9 | 1 |
| 173 | 16 | 16 | 8 | 8 | 9 | 1 |
| 174 | 16 | | 8 | | 8 | 0 |
| 175 | 15 | 15 | 8 | 8 | 10 | 2 |
| 176 | 15 | | 8 | | 9 | 1 |
| 177 | 15 | 15 | 7 | 7 | 8 | 1 |
| 178 | 15 | | 7 | | 7 | 0 |
| 179 | 15 | 15 | 8 | 8 | 6 | -2 |
| 180 | 16 | 16 | 9 | 9 | 13 | 4 |
| 181 | 16 | | 9 | | 12 | 3 |
| 182 | 16 | | 9 | | 10 | 1 |
| 183 | 16 | | 9 | | 9 | 0 |
| 184 | 15 | 15 | 7 | 7 | 11 | 4 |
| 185 | 15 | | 7 | | 10 | 3 |
| 186 | 15 | | 7 | | 9 | 2 |
| 187 | 16 | 16 | 7 | 7 | 8 | 1 |
| 188 | 15 | 15 | 7 | 7 | 7 | 0 |
| 189 | 15 | 15 | 7 | 7 | 11 | 4 |
| 190 | 15 | | 7 | | 10 | 3 |
| 191 | 15 | | 7 | | 9 | 2 |
| 192 | 16 | 16 | 8 | 8 | 9 | 1 |
| 193 | 15 | 15 | 6 | 6 | 7 | 1 |
| 194 | 15 | | 6 | | 6 | 0 |
| 195 | 16 | 16 | 9 | 9 | 9 | 0 |
| 196 | 16 | 16 | 9 | 9 | 8 | -1 |
| 197 | 16 | 16 | 8 | 8 | 7 | -1 |
| 198 | 15 | 15 | 6 | 6 | 7 | 1 |
| 199 | 15 | | 6 | | 6 | 0 |
| 200 | 15 | | 6 | | 5 | -1 |
| 201 | 15 | 15 | 6 | 6 | 9 | 3 |
| 202 | 15 | | 6 | | 8 | 2 |
| 203 | 15 | | 6 | | 7 | 1 |
| 204 | 15 | | 6 | | 6 | 0 |
| 205 | 16 | 16 | 8 | 8 | 12 | 4 |
| 206 | 16 | | 8 | | 11 | 3 |
| 207 | 16 | | 8 | | 10 | 2 |
| 208 | 16 | 16 | 7 | 7 | 9 | 2 |
| 209 | 16 | | 7 | | 7 | 0 |
| 210 | 15 | 15 | 8 | 8 | 6 | -2 |
| 211 | 14 | 14 | 8 | 8 | 7 | -1 |
| 212 | 14 | | 8 | | 6 | -2 |
| 213 | 16 | 16 | 8 | 8 | 12 | 4 |
| 214 | 16 | | 8 | | 11 | 3 |
| 215 | 16 | 16 | 9 | 9 | 10 | 1 |
| 216 | 16 | | 9 | | 9 | 0 |
| 217 | 17 | 17 | 8 | 8 | 9 | 1 |
| 218 | 17 | | 8 | | 8 | 0 |
| 219 | 17 | | 8 | | 7 | -1 |
| 220 | 16 | 16 | 8 | 8 | 10 | 2 |
| 221 | 15 | 15 | 7 | 7 | 9 | 2 |
| 222 | 15 | | 7 | | 8 | 1 |
| 223 | 15 | 15 | 6 | 6 | 7 | 1 |
| 224 | 15 | | 6 | | 5 | -1 |
| 225 | 15 | 15 | 8 | 8 | 12 | 4 |
| 226 | 15 | | 8 | | 11 | 3 |
| 227 | 15 | | 8 | | 10 | 2 |
| 228 | 15 | | 8 | | 9 | 1 |
| 229 | 16 | 16 | 10 | 10 | 9 | -1 |
| 230 | 15 | 15 | 9 | 9 | 10 | 1 |
| 231 | 14 | 14 | 7 | 7 | 7 | 0 |
| 232 | 14 | 14 | 8 | 8 | 6 | -2 |
| 233 | 16 | 16 | 8 | 8 | 8 | 0 |
| 234 | 16 | 16 | 9 | 9 | 9 | 0 |
| 235 | 16 | | 9 | | 8 | -1 |
| 236 | 16 | | 9 | | 7 | -2 |
| 237 | 15 | 15 | 7 | 7 | 9 | 2 |
| 238 | 15 | | 7 | | 8 | 1 |
| 239 | 15 | | 7 | | 6 | -1 |
| 240 | 16 | 16 | 8 | 8 | 9 | 1 |
| 241 | 16 | | 8 | | 8 | 0 |

## V. Separation distance between the positions of the richest Co and dislocations for the 0.4-sample

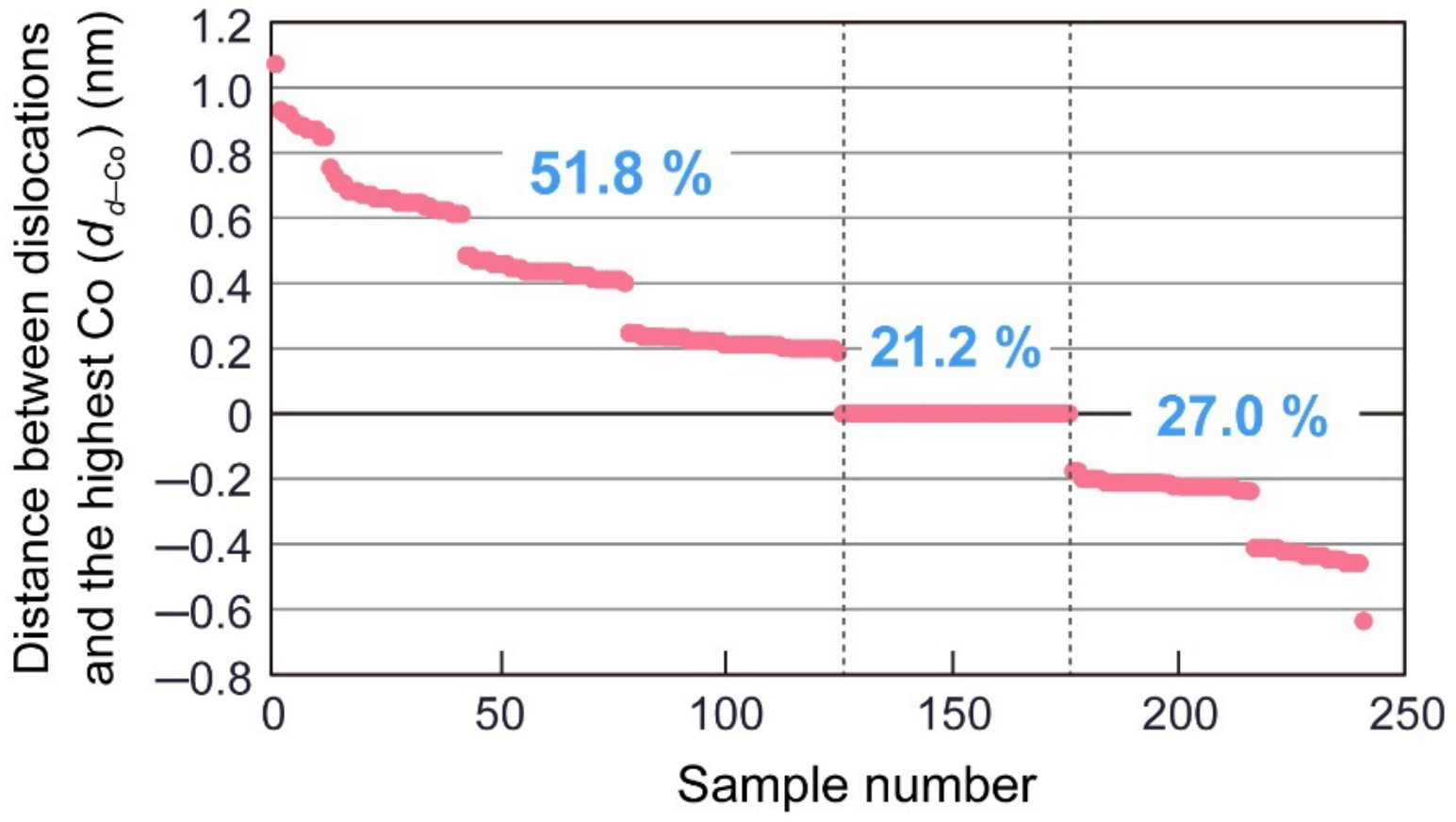

FIG. S5. Separation distance $d_{d\text{-}Co}$ between the positions of the richest Co and dislocations for the 0.4-sample. This figure corresponds to Fig. 4(e) in the main text. The numerical data are listed in Table S4.

## VI. Deposition system geometry and sputtering parameters

The multilayers were deposited by magnetron sputtering using a system equipped with 2-inch diameter targets. The target-to-substrate distance was 110 mm, and the substrate was positioned at normal incidence with respect to the sputtering guns. All depositions were performed at room temperature without active substrate heating; a slight increase in substrate temperature due to plasma exposure during deposition cannot be excluded. The substrate was electrically grounded throughout all depositions.

Pt and Co layers were deposited by DC magnetron sputtering, while Ta layers were deposited by RF magnetron sputtering. For RF sputtering of Ta, the reported voltage corresponds to the self-bias DC voltage ($V_{dc}$) generated during RF operation. The applied powers and corresponding target voltages for each layer at each argon pressure are summarized in Table S7.

| Ar pressure (Pa) | Pt power (W) | Pt voltage (V) | Co power (W) | Co voltage (V) | Ta power (W) | Ta $V_{dc}$ (kV) |
|---|---|---|---|---|---|---|
| 0.4 | 15 | -327 | 25 | -326 | 60 | -0.10 |
| 1.0 | 15 | -299 | 25 | -277 | 60 | -0.09 |
| 1.5 | 15 | -286 | 25 | -268 | 60 | -0.08 |
| 2.0 | 15 | -282 | 25 | -259 | 60 | -0.08 |

Table S7. Sputtering power and target voltage for each layer at each argon pressure. For Ta (RF sputtering), the voltage corresponds to the self-bias $V_{dc}$.

These parameters show that increasing the argon pressure results in a systematic decrease in the target voltage at constant power for both Pt and Co, consistent with the well-known reduction in mean free path and modification of the sputtered atom energy distribution at higher pressures. This pressure-dependent change in adatom kinetics is expected to influence the film microstructure, including grain size, defect density, and intermixing, as discussed in the main text.

**References**

Refs. S1–S9 below are identical to Refs. 42, 45, 44, 46, 48–52, respectively, in the main text.

[S1] A. Thiaville, S. Rohart, É. Jué, V. Cros, and A. Fert, "Dynamics of Dzyaloshinskii domain walls in ultrathin magnetic films," EPL **100**, 57002 (2012).

[S2] A. Vansteenkiste, J. Leliaert, M. Dvornik, M. Helsen, F. Garcia-Sanchez, B. Van Waeyenberge, "The design and verification of MuMax3, " AIP Adv. **4**, 107133 (2014).

[S3] W. Legrand, J. Y. Chauleau, D. Maccariello, N. Reyren, S. Collin, K. Bouzehouane, N. Jaouen, V. Cros, and A. Fert, "Hybrid chiral domain walls and skyrmions in magnetic multilayers," Sci. Adv. **4**, eaat0415 (2018).

[S4] N. S. Gusev, A. V. Sadovnikov, S. A. Nikitov, M. V. Sapozhnikov, and O. G. Udalov, "Manipulation of the Dzyaloshinskii–Moriya interaction in Co/Pt multilayers with strain," Phys. Rev. Lett. **124**, 157202 (2020).

[S5] J. W. Lee, Y.-W. Oh, S.-Y. Park, A. I. Figueroa, G. van der Laan, G. Go, K.-J. Lee, and B.-G. Park, "Enhanced spin-orbit torque by engineering Pt resistivity in Pt/Co/$AlO_x$ structures," Phys. Rev. B **96**, 064405 (2017).

[S6] R. Wang, Z. Xiao, H. Liu, Z. Quan, X. Zhang, M. Wang, M. Wu, and X. Xu, "Enhancement of perpendicular magnetic anisotropy and spin-orbit torque in Ta/Pt/Co/Ta multi-layered heterostructures through interfacial diffusion," Appl. Phys. Lett. **114**, 042404 (2019).

[S7] D. K. Ojha, R. Chatterjee, Y.-L. Lin, Y.-H. Wu, P.-W. Chen, and Y.-C. Tseng, "Spin-torque efficiency enhanced in sputtered topological insulator by interface engineering," J. Magn. Magn. Mater. **572**, 170638 (2023).

[S8] L. Chen, K. Zhang, B. Li, B. Hong, W. Huang, Y. He, X. Feng, Z. Zhang, K. Lin, W. Zhao, and Y. Zhang, "Engineering symmetry breaking enables efficient bulk spin-orbit torque-driven perpendicular magnetization switching," Adv. Funct. Mater. **34**, 2308823 (2023).

[S9] B. T. Thole and G. van der Laan, "Branching ratio in x-ray absorption spectroscopy," Phys. Rev. B **38**, 3158 (1988).